\documentclass[12pt]{article}
\pdfoutput=1
\usepackage{tikz}
\usepackage{mathrsfs}
\usepackage{amsmath,amsfonts}
\usepackage{latexsym}
\usepackage{amsthm}
\usepackage{amssymb}
\usepackage{subfig}
\usepackage{xfrac}
\usepackage{enumerate}
\usepackage{adjustbox}
\usepackage{extarrows}
\usepackage{cite}
\definecolor{darkblue}{cmyk}{0.9,0.9,0,0}
\usepackage[colorlinks=true,linkcolor=darkblue,citecolor=darkblue,urlcolor=darkblue]{hyperref}
\usepackage{wasysym}
\usepackage{varioref}
\usepackage[english]{babel}
\usepackage{tikz}
\usepackage{booktabs}

\usepackage[font={small}]{caption}

\usepackage{microtype}

\newcommand\smallO{
  \mathchoice
    {{\scriptstyle\mathcal{O}}}
    {{\scriptstyle\mathcal{O}}}
    {{\scriptscriptstyle\mathcal{O}}}
    {\scalebox{.7}{$\scriptscriptstyle\mathcal{O}$}}
  }

\def\braket#1{\langle #1 \rangle}

\newcommand{\zb}{\bar{z}}
\newcommand{\HPL}[2]{H_{#1}\left(#2\right)}

\begin{document}

\thispagestyle{empty}

\renewcommand{\thefootnote}{\fnsymbol{footnote}}
\setcounter{page}{1}
\setcounter{footnote}{0}
\setcounter{figure}{0}


\noindent

\hfill
\begin{minipage}[t]{35mm}
\begin{flushright}
UUITP-45/20 
\end{flushright}
\end{minipage}

\vspace{1.0cm}
\numberwithin{equation}{section}
\begin{center}
{\Large\textbf{\mathversion{bold}
All loop structures in Supergravity Amplitudes on $AdS_5 \times S^5$ from CFT\\
}\par}

\vspace{1.0cm}

\textrm{Agnese Bissi\textsuperscript{1}, Giulia Fardelli\textsuperscript{1}, Alessandro Georgoudis\textsuperscript{1,2,3}}
\\ \vspace{1.2cm}
\footnotesize{\textit{
\textsuperscript{1}Department of Physics and Astronomy, Uppsala University
Box 516, SE-751 20 Uppsala, Sweden\\
\textsuperscript{2} Institut de Physique Th\'eorique, CEA, CNRS, Universit\'e Paris-Saclay, F-91191 Gif-sur-Yvette cedex, France \\
\textsuperscript{3}Laboratoire de physique de l'Ecole normale sup\'erieure, ENS, Universit\'e PSL, CNRS, Sorbonne Universit\'e, Universit\'e Paris-Diderot, Sorbonne Paris Cit\'e, 24 rue Lhomond, 75005 Paris, France
}  
\vspace{4mm}
}

\end{center}

\begin{abstract}
We compute a set of structures which appear in the four-point function of protected operators of dimension two in $\mathcal{N}=4$ Super Yang Mills with $SU(N)$ gauge group, at any order in a large $N$ expansion. They are determined only by leading order CFT data. By focusing on a specific limit, we make connection with the dual supergravity amplitude in flat space, where such structures correspond to iterated $s$-cuts. We make several checks and we conjecture that the same interpretation holds for supergravity amplitudes on $AdS_5 \times S^5$. 
\end{abstract}
\noindent
\setcounter{page}{1}
\renewcommand{\thefootnote}{\arabic{footnote}}
\setcounter{footnote}{0}

\setcounter{tocdepth}{2}

\newpage

\parskip 5pt plus 1pt   \jot = 1.5ex

\newpage

\parskip 5pt plus 1pt   \jot = 1.5ex

\tableofcontents

\section{Introduction and summary}
Since the advent of the AdS/CFT correspondence a lot of progress has been achieved in the computation of observables, in particular correlation functions of local and non local operators or equivalently scattering amplitudes of the dual string states. Despite the huge effort, there is not yet a working strategy to compute the above-mentioned observables in four dimensional $\mathcal{N}$=4 Super Yang Mills (SYM) with gauge group $SU(N)$, for generic values of the coupling constant $g_{YM}$ and of the number of colours $N$. In this paper we make use of the holographic dictionary and of the symmetry group of $\mathcal{N}$=4 SYM to find structures which appear in the four-point correlation function of protected operators of dimension two, for large 't Hooft coupling $\lambda=g_{YM}^2 N$ and at all loops  in the large $N$ expansion. In this particular regime, often called supergravity limit, we are probing the four-point graviton amplitude in the two derivative supergravity effective action and at any genus. This limit makes clear the fact that traditional techniques such as diagrammatic Witten diagram computations are generically very hard or even impossible to be carried out. In the last years there has been a considerable progress in computing such observables combining the methods of the analytic bootstrap, localization, and integrability \cite{Aharony:2016dwx,Chester:2020vyz,Chester:2020dja,Chester:2019pvm,Chester:2019jas,Aprile:2019rep,Aprile:2020uxk,Drummond:2020uni,Komatsu:2020sag,Bissi:2020wtv,Alday:2017vkk,Aprile:2018efk,Alday:2018pdi,Alday:2019nin,Alday:2018kkw,Binder:2019jwn,Drummond:2019hel,Bargheer:2019exp,Drummond:2020dwr}. In particular it has been possible to compute the four-point function of half-BPS operators of protected dimension two  $\mathcal{O}_2$, in a large $N$ expansion up to order $\frac{1}{N^{4}}$ and as an expansion in inverse powers of $\lambda$ . In the supergravity regime, the intermediate states in the operator product expansion (OPE) of $\mathcal{O}_2 \times \mathcal{O}_2$ are double trace operators of the schematic form $[\mathcal{O}_2\mathcal{O}_2]_{n,\ell}\sim \mathcal{O}_2 \Box^n \partial^{\ell}\mathcal{O}_2 $ with dimension $\Delta_{n,\ell}=4+2n+\ell+\frac{\gamma^{(1)}_{n, \ell}}{N^2}+\frac{\gamma^{(2)}_{n, \ell}}{N^4}+\cdots$. To compute the anomalous dimension $\gamma^{(2)}_{n, \ell}$ it has been crucial to notice that at leading order there is a degeneracy among the different double trace operators of the form $[\mathcal{O}_p\mathcal{O}_p]_{n,\ell}$ which have bare dimension $2p+2n+\ell$. Up to order $\frac{1}{N^2}$ the mixing has been solved, which means that both the square of the three-point functions $a^{(0)}_{n,\ell,I}\sim \langle \mathcal{O}_2 \mathcal{O}_2 O_{n,\ell,I} \rangle $ and the anomalous dimensions $\gamma^{(1)}_{n, \ell,I}$ have been computed, where $\mathcal{O}_{n,\ell,I}$ denotes the eigenstate of the dilatation generator which is a specific linear combination of the double trace operators $[\mathcal{O}_p\mathcal{O}_p]_{n, \ell}$ \cite{Alday:2017xua,Aprile:2017bgs,Aprile:2017xsp}. In addition, up to order $\frac{1}{N^4}$ the double discontinuity (dDisc) of the four-point function is completely determined by $a^{(0)}_{n,\ell,I}$ and $\gamma^{(1)}_{n,\ell,I}$, thus making possible to compute the four-point function at this order, also including corrections in inverse power of $\lambda$\footnote{At order $\frac{1}{N^4}$ there is a contact term like ambiguity, which only influences the CFT data corresponding to the intermediate scalar operators in the OPE. }. One interesting comment to make is that in all studied cases, there has been a matching between a specific limit in the correlation function, which is the one probing  the bulk point singularity, and the flat space limit of the amplitude \cite{Alday:2017vkk}.

Going higher in the $\frac{1}{N}$ expansion has several difficulties, both technical and conceptual. The first one is that also at leading order in the 't Hooft coupling $\lambda$, there are additional operators in the OPE of $\mathcal{O}_2 \times \mathcal{O}_2$, in particular there are triple trace operators of the schematic form $[\mathcal{O}_{p_1} \mathcal{O}_{p_2}\mathcal{O}_{p_3}]$ which can potentially mix among themselves and with the double trace ones. Another obstacle resides in the fact that the dDisc gets contributions from other terms in the conformal blocks expansion which cannot be expressed in terms only of tree level and one loop CFT data. Despite these obstacles, at any given order $\frac{1}{N^{2\kappa}}$ the four-point function $\mathcal{G}^{(\kappa)}(z,\bar{z})$ contains a piece which can be fully reconstructed once we know the leading order data
\begin{equation}
\log^{\kappa}(z \zb) \sum_{I, n, \ell}\frac{a^{(0)}_{n,\ell,I} \left(\gamma^{(1)}_{n,\ell,I}\right)^{\kappa}}{2^{\kappa}\kappa!} (z \zb)^{n+2} g_{8+2n,\ell}(z,\zb) \subset \mathcal{G}^{(\kappa)}(z,\bar{z}) \, ,
\end{equation}
where the cross ratios enter as $U=z \bar{z}$ and $V=(1-z)(1-\bar{z})$, $g_{n,\ell}(z,\zb)$ are the four dimensional conformal blocks and $\mathcal{G}^{(\kappa)}(z,\bar{z})$ is the four-point function at order $\frac{1}{N^{2\kappa}}$\footnote{The precise definitions will be given later in the text.}. The main idea of our work is to understand to which part of the dual amplitude this piece corresponds to at any given $\kappa$ order. To get to this relationship we study the bulk point limit and put it in relation with the flat space limit of the corresponding amplitude, by comparing the discontinuities of the two observables. In a previous work \cite{Bissi:2020wtv} we have unravelled the case of $\kappa=3$ and we have seen that the double cut of the amplitude exactly reproduces the flat space limit of the CFT and only conjectured that the same behaviour would persist at any loop order.
In this paper we have substantiated this observation and we have constructed the iterated $s$-cut to all loops. We have found that the bulk point limit of the double discontinuity of the leading logarithmic term in $\mathcal{G}^{(\kappa)}$, once logarithms are factored out, is reproduced by the iterated $s$-channel cut of the corresponding $\kappa$ rungs ladder diagram. Pictorially this reads

\begin{align}
  \sum_{n,\ell} \sum_{I=1}^{n+1} (z \zb)^{n+2} a^{(0)}_{n,\ell, I}(\gamma_{n,l, I}^{(1)})^{\kappa}g_{2n+8, \ell}(z, \zb) \xleftrightarrow{\text{flat space}}\begin{tikzpicture}[baseline={([yshift=-1.1ex]current bounding box.center)},scale=0.25,node/.style={draw,shape=circle,fill=black,scale=0.4}]
   \draw [thick] (0,0) -- (6,0)--(6,3)--(0,3)--(0,0);
    \draw [thick] (3,0) -- (3,3);
    \draw [thick] (0,0) -- (-1,-1);
    \draw [thick] (0,3) -- (-1,4);
    \draw [thick] (6,0) -- (6.8,0);
    \draw [thick] (6,3) -- (6.8,3);
    \draw [thick,dashed] (7.2,0) -- (9.8,0);
    \draw [thick,dashed] (7.2,3) -- (9.8,3);
    \draw [thick] (9.3,0) -- (13,0);
    \draw [thick] (9.3,3) -- (13,3);
    \draw [thick] (10,0) -- (10,3);
    \draw [thick] (13,0) -- (13,3);
    \draw [thick] (13,0) -- (14,-1);
    \draw [thick] (13,3) -- (14,4);
    \draw [dashed,thick,red] (1.5,4)--(1.5,-1);
    \draw [dashed,thick,red] (4.5,4)--(4.5,-1);
    \draw [dashed,thick,red] (11.5,4)--(11.5,-1);
  \end{tikzpicture} \, .
\end{align}

This allowed us to notice some interesting features on how the
unitarity structure translates in the flat-space limit. First of all when studying the discontinuity of the amplitude, it is evident the appearance of multi-particle states in the cuts one has to construct, thus suggesting the need to include higher and higher trace operators in the OPE in the large $N$ expansion of the correlator. At two loops we have noticed that the structure of the double
particle contribution in the amplitude is fixed by its iterated $s$-cut
construction contrary to the correlator where the knowledge of the
$\log^2U$ part can not be fully reconstructed. The most important remark that we would like to make at this point is the fact that we expect such $s$-channel cutting holds also in the full-fledged amplitudes in the $AdS_5 \times S^5$ background, on the line also of \cite{Meltzer:2019nbs, Meltzer:2020qbr}.\\
An interesting interpretation of the iterated cut depicted before, which formally can be written as the corresponding Feynman integral with insertions of delta functions for each cut propagator as in Eq.~\eqref{iteratedSch}, can be given when considering the theory of Landau singularities  \cite{Eden:1966dnq}. In this context in fact, this quantity measures the discontinuity of the \textit{lower-order} Landau singularity associated to $\kappa$ contractions. A  \textit{lower-order} Landau singularity is defined as the \textit{leading singularity} of a contracted graph, a Feynman diagram where all the internal legs, that are not on-shell, are pinched together to a point. In our specific case,
\begin{align*}
\begin{array}{c}
\text{\textit{lower-order} singularity associated} \\  
\text{to $\kappa$ contractions }
\end{array} 
\quad \Leftrightarrow \quad
\begin{array}{c}
 \text{ \textit{leading} singularity of } \\  \vspace{-0.25cm}
 \\ \begin{tikzpicture}[baseline={([yshift=-1.1ex]current bounding box.center)},scale=0.35,node/.style={draw,shape=circle,fill=black,scale=0.4}]
  \def\ys{0};
     \draw [thick] (1.5-1,2-\ys) ellipse (1.5cm and 1cm);
   \draw [thick] (4.5-1,2-\ys) ellipse (1.5cm and 1cm);
   \draw [thick] (10.5-1,2-\ys) ellipse (1.5cm and 1cm);
   \draw [thick,densely dashed] (7.5-1,2) ellipse (1.5 cm and 1 cm);
   \draw [thick] (9-1,2) arc(0:40:1.5cm and 1cm);
   \draw [thick] (6-1,2) arc(180:230:1.5cm and 1cm);
   \draw [thick] (6-1,2) arc(180:140:1.5cm and 1cm);
   \draw [thick] (9-1,2) arc(0:-30:1.5cm and 1cm);
    \draw [thick] (0-1,2-\ys) -- (-1-1,1-\ys);
    \draw [thick] (0-1,2-\ys) -- (-1-1,3-\ys);
    \draw [thick] (12-1,2-\ys) -- (13-1,1-\ys);
    \draw [thick] (12-1,2-\ys) -- (13-1,3-\ys);
    \node at (-1,2-\ys) [circle,fill,inner sep=1.8pt]{};
    \node at (2,2-\ys) [circle,fill,inner sep=1.8pt]{};
     \node at (5,2-\ys) [circle,fill,inner sep=1.8pt]{};
      \node at (8,2-\ys) [circle,fill,inner sep=1.8pt]{};
       \node at (11,2-\ys) [circle,fill,inner sep=1.8pt]{};
  \end{tikzpicture}
  \end{array}
\end{align*}
Finally from our analysis in Mellin space, it emerges even more clearly the interplay between unitarity cuts of  flat space amplitudes and CFT correlator with a certain logarithmic behaviour in the small $U$ and $V$ expansion. In particular we conjecture that the Mellin amplitude corresponding to the leading logarithmic terms at a given $\kappa$ order  and in the large AdS radius limit  can be derived from a \textit{maximal cut} of the corresponding ladder diagram. 

In summary,in this paper we study a specific contribution of the four point correlator, which is proportional to $\log^{\kappa} U$, for $\kappa=3$ we computed the full function, and for higher $\kappa$ we give an explicit method to compute it. In the specific limit that we compare to the the flat space limit of the amplitude, we computed this function to several values of $\kappa$, which label the loop order.  Given the evidences of the explicit examples, we conjecture that at any loop order the leading logs of the four-point function, completely fixed by known tree level data, reproduce in the flat space limit iterated $s$-channel cuts of the dual graviton amplitude. Since only planar ladders  admit exactly $(\kappa-1)$ two-particle cuts at $(G_N)^{\kappa}$, we restrict our discussion just to these diagrams.  Moreover a similar analysis, performed in Mellin space, enables us to establish a connection between the residues of $(s-2m)^{-(\kappa-1)}(t-2n)^{-1}$ terms and  maximal cuts. \\
Due to the appearance of higher trace operators and the unsolved mixing problem, we do not go  beyond $\log^{\kappa} U$ contributions. Nonetheless in the specific $\kappa=3$ example we make clear how triple trace operators enter the discussion and how to disentangle their contribution at the level of the amplitude. 

\begin{figure}[h]
\begin{center}
\includegraphics[width=\textwidth]{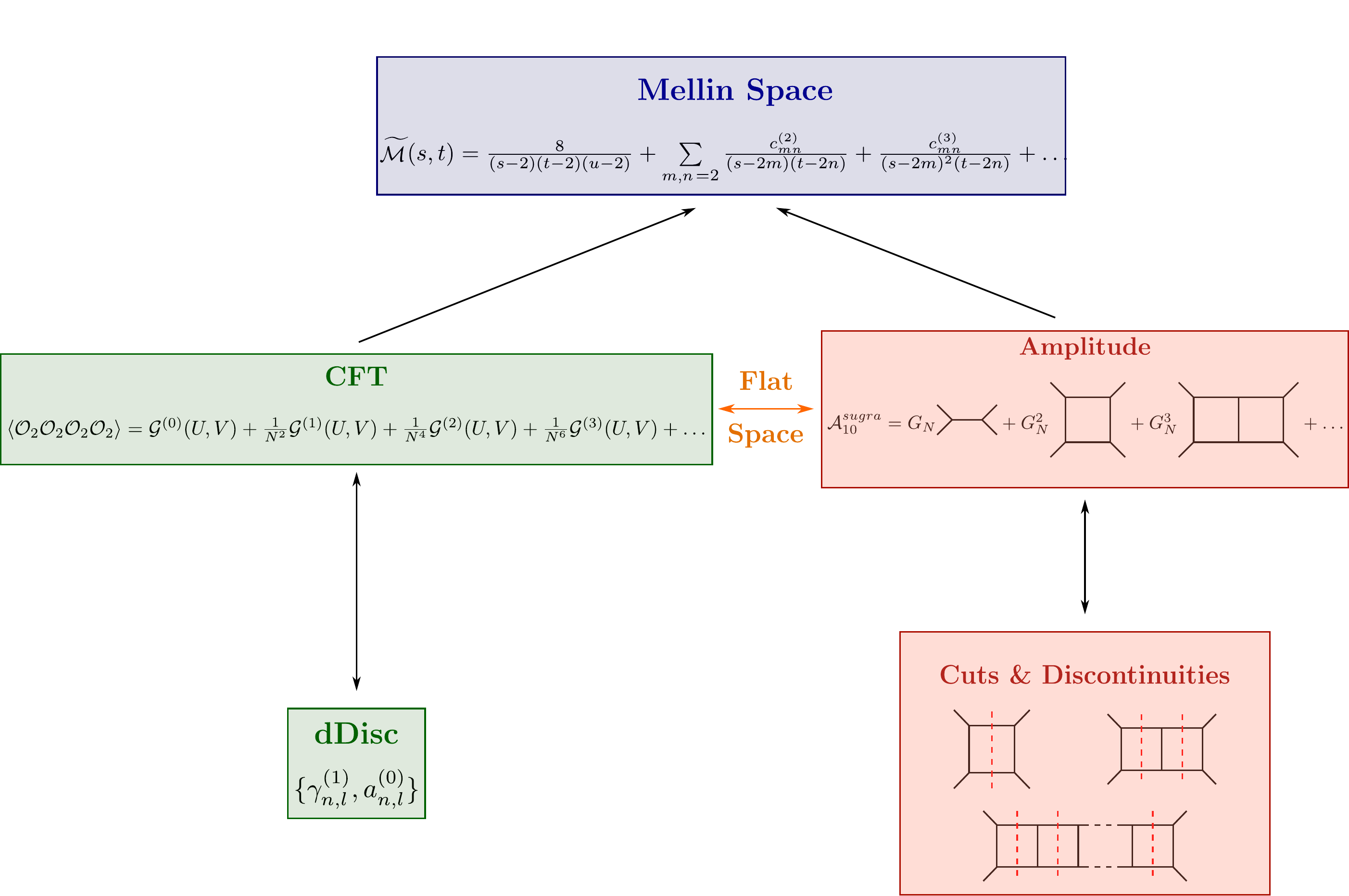}
\caption[]{Visual idea of the working strategy we will be following in the paper. \footnotemark}
\end{center}
\end{figure}
\newpage
\footnotetext{Notice that on the amplitude side we are not explicitly reporting the disconnected part, which is dual to the $\mathcal{G}^{(0)}$ term in the CFT correlator.}
 {\textbf{Open questions}}
 \begin{itemize}
 \item {\bf{Include stringy corrections}}: While we worked in a regime in which the 't Hooft coupling constant is strictly infinite, it would be interesting to add the full expansion in inverse powers of $\lambda$. This would allow including quartic vertices in the amplitudes, along the lines of \cite{Alday:2018pdi}. We believe that the same interpretation in terms of $s$-cuts holds in this case as well. Going beyond the supergravity limit would also require the introduction of single trace operators in the OPE, in addition to the double trace and higher trace operators which are already taken into account. Enlarging the set of operators will make possible to go away from the strong coupling regime and probe finite coupling. 
 \item {\bf{Resum iterated $(\kappa-1)$-cut}}: Given the simplicity and the iterative form of  multiple $s$-cuts for ladder diagrams, it would be interesting to resum this contribution to all orders and explore possible physical interpretation.  {This seems reminiscent of leading Regge behaviour of  four-point amplitudes.  A similar behaviour has been extensively studied in QCD and solved using a Wilson line approach \cite{Korchemsky:1993hr}.}
 \item {\bf{Complement our results with string theory computations}}: Low energy expansion of string theory amplitudes has a revival in the last years. In particular old fashioned methods have been complemented with insights from modular graph forms and the closed/open string amplitudes relations, along the lines of \cite{DHoker:2015wxz, DHoker:2016mwo, DHoker:2017pvk, DHoker:2018mys, DHoker:2020uid, Gerken:2020yii, Gerken:2020xfv}. In the spirit of computing string amplitudes, at any value of the central charge and of the 't Hooft coupling, the synergy of these methods can be helpful and can provide insights that can be useful in both methods. 
  \end{itemize}
The paper is organised as follows. In Section~\ref{sec:FourPt} we study the $\braket{\mathcal{O}_2\mathcal{O}_2\mathcal{O}_2\mathcal{O}_2}$ correlation function in the large $N$ expansion focusing on its leading logarithmic terms in the flat space limit. The corresponding quantity on the gravity side, namely the supergravity four-graviton amplitude in $\mathbb{R}^{10}$, is at the centre of Section~\ref{sec:Amplitude}. Here we very briefly collect the main findings for the two-loop example and, starting from them, we extend the results to all loops introducing the idea of iterated $s$-channel unitarity cuts. In Section~\ref{sec:Mellin} we address similar problems, but now rephrased in Mellin space. Some Appendices  collect the lengthier computations and other technical details, in particular in Appendix \ref{App:2loop} a full analysis of the two-loop correlator is given, complementing the discussion of our previous work \cite{Bissi:2020wtv}. 
\section{Four-point function}\label{sec:FourPt}
In this paper we focus on the study of the four-point correlation function of dimension two protected operators in $\mathcal{N}=4$ SYM in four dimensions with gauge group $SU(N)$. These are the lowest dimension half-BPS operators $\mathcal{O}_p$: superconformal primary operators with protected dimension $\Delta=p$, scalars under the Lorentz group and transforming in the  $[0,p,0]$ representation of the R-symmetry group $SU(4)_R \simeq SO(6)_R$. In terms of the six scalar fields $\Phi^{i}$ ($i=1, \ldots, 6$)  of $\mathcal{N}=4$  theory, they can be written  as \textit{single trace} operators of the form
\begin{align}
\label{half-BPS}
\mathcal{O}_p(x, y)=y_{i_1}\, \ldots\, y_{i_p} \text{Tr}\,\Phi(x)^{\{i_1}\, \ldots \,  \Phi(x)^{i_p\}} \qquad\qquad  p=2, 3, \, \ldots 
\end{align}
Here we have introduced the $SO(6)$ null vectors $y_i$, $y \cdot y=0$, such that the $\mathcal{O}_p$'s manifestly transform as  symmetric traceless tensors. For later use, multi-trace operators will be given by normal product of single-trace ones taken such that they transform in a symmetric and traceless way. \\
Through the AdS/CFT dictionary half-BPS operators, or to be more precise a combination of \eqref{half-BPS} and multi-trace operators $\mathcal{O}(1/N)$ suppressed \cite{Aprile:2018efk,Alday:2019nin,Aprile:2020uxk}, are mapped to \textit{single particle} states in AdS.
\subsection{$\braket{\mathcal{O}_2 \mathcal{O}_2 \mathcal{O}_2 \mathcal{O}_2}$ correlator} 
The $\mathcal{O}_2$ operator at the centre of our investigation represents the superconformal primary of the stress tensor multiplet: it is a scalar with $\Delta=2$ and it transforms under the $\mathbf{20^{\prime}}$ representation of $SU(4)_R$. On the gravity side it maps directly to a single particle state,  the scalar belonging to the graviton multiplet in type IIB closed strings in AdS$_5 \times$S$^5$ background, while the $\mathcal{O}_p$'s with $p>2$ correspond to Kaluza Klein modes.

Conformal invariance and supersymmetry highly constrain the form of its four-point function \cite{Nirschl:2004pa, Dolan:2004mu}:
\begin{align}
\braket{\mathcal{O}_2(x_1, y_1)\mathcal{O}_2(x_2, y_2)\mathcal{O}_2(x_3, y_3)\mathcal{O}_2(x_4, y_4)}=\frac{(y_1 \cdot y_2)^2(y_3 \cdot y_4)^2}{x_{12}^4 x_{34}^4}\sum_{\mathcal{R}}Y^{\mathcal{R}}(\sigma, \tau) \mathcal{G}^{\mathcal{R}}(U, V) \, ,
\end{align}
with $x_{i j}=x_i-x_j $ and where we have defined respectively the position and polarization space cross-ratios as 
\begin{align}
U&=\frac{x_{12}^2 x_{34}^2}{x_{13}^2 x_{24}^2}=z \zb \, , && V=\frac{x_{14}^2 x_{23}^2}{x_{13}^2 x_{24}^2}=(1-z)(1-\zb) \, ,\\ \label{crossRatios}
\sigma &= \frac{(y_1 \cdot y_3)(y_2 \cdot y_4)}{(y_1 \cdot y_2)(y_3 \cdot y_4)}\, , &&\tau =\frac{(y_1 \cdot y_4)(y_2 \cdot y_3)}{(y_1 \cdot y_2)(y_3 \cdot y_4)} \, .
\end{align}
$Y^{\mathcal{R}}(\sigma, \tau)$ are $SO(6)_R$ harmonics and the sum runs over all possible representations $\mathcal{R} \in [0,2,0] \otimes [0,2,0]=[0,0,0]\oplus[0,2,0]\oplus[0,4,0]\oplus[1,0,1]\oplus[1,2,1]\oplus[2,0,2]$. However thanks to superconformal Ward identities \cite{Beem:2016wfs}, one can actually reduce the study of the $\mathcal{G}^{\mathcal{R}}$ to only one function, namely $\mathcal{G}(U,V)\equiv \frac{\mathcal{G}^{\mathbf{105}}}{U^2}$ with $\mathbf{105}=[0,4,0]$. This function should fulfil the crossing symmetry condition
\begin{align}\label{crossingG}
V^2 \mathcal{G}(U,V)-U^2 \mathcal{G}(V, U)+(U^2-V^2)+\frac{U-V}{c}=0 \, ,
\end{align}
where $c$ represents the central charge. From the function $\mathcal{G}(U,V)$ one can further disentangle the contributions from protected and non protected operators in the Operator Product Expansion (OPE), such that
\begin{align}
\mathcal{G}(U,V)&=\mathcal{G}^{short}(U,V)+\mathcal{H}(U,V) \, ,
\end{align}
where $\mathcal{G}^{short}(U,V)$ is fully known and $c^{-1}$ exact \cite{Beem:2016wfs}, and receives only contributions from the protected sector consisting of short and semi-short multiplets. The interacting and coupling dependent part $\mathcal{H}(U,V)$ admits an expansion in superconformal blocks of the form
\begin{align}
\label{Hexpanion}
\mathcal{H}(z,\zb)&=\sum_{\tau, \ell} a_{\tau,\ell}(z \zb)^{\tau/2} g_{\tau+4,\ell}(z,\zb)\, .
\end{align}
The sum runs over non protected long primaries with twist, dimension minus the spin, $\tau$, even spin $\ell$ and which are singlet under $SU(4)_R$. $a_{\tau,\ell}$ indicates the square of the OPE coefficient and the superconformal blocks are defined as
\begin{equation}
g_{\tau, \ell}=\frac{z^{\ell+1}k_{\tau+2\ell}(z)k_{\tau-2}(\zb)-\zb^{\ell+1}k_{\tau+2l}(\zb)k_{\tau-2}(z)}{z-\zb} \, ,
\end{equation}
\begin{equation}
k_{\beta}(x)={}_2F_1(\frac{\beta}{2}\frac{\beta}{2},\beta;x)\, .
\end{equation}
We study the four-point function in the supergravity regime, i.e. when the 't Hooft coupling $\lambda=g_{YM}^2 N$ is taken to infinity, and as an expansion at large $N$ or equivalently at large $c=\frac{N^2-1}{4}$:
\begin{align}
\label{eq:c-expansion}
\mathcal{H}(z,\zb)=&\mathcal{H}^{(0)}(z,\zb)+\frac{1}{c}\mathcal{H}^{(1)}(z,\zb)+\frac{1}{c^2}\mathcal{H}^{(2)}(z,\zb)+\frac{1}{c^3}\mathcal{H}^{(3)}(z,\zb)+\dots \, ,
\end{align}
where $\mathcal{H}^{(1)}(z,\zb)=-(z \zb)^2\bar{D}_{2,4,2,2}(z, \zb)$ \cite{Arutyunov:2000py, Dolan:2001tt}. The crossing condition~\eqref{crossingG} now takes two different forms, for $\kappa=0,\, 1$:
\begin{align}\nonumber
&(1-z)^2(1-\zb)^2 \mathcal{G}^{short}(z, \zb)-z^2 \zb^2 \mathcal{G}^{short} (1-z, 1-\zb)+( z^2 \zb^2-(1-z)^2(1-\zb)^2)=\\&\qquad -\frac{z \zb-(1-z)(1-\zb)}{c}+
\sum_{\kappa=0}^1 \left( -(1-z)^2(1-\zb)^2\frac{\mathcal{H}^{(\kappa)}(z, \zb)}{c^{\kappa}}+z^2 \zb^2 \frac{\mathcal{H}^{(\kappa)}(1-z, 1-\zb)}{c^{\kappa}} \right) \, ,
\end{align}
while for $\kappa \geq 2$,
\begin{align} 
\label{eq:crossing}
&(1-z)^2(1-\zb)^2\mathcal{H}^{(\kappa)}(z,\zb)=z^2 \zb^2
\mathcal{H}^{(\kappa)}(1-z,1-\zb) \, .
\end{align}
In the supergravity limit and up to order $c^{-1}$ the only long operators exchanged in the OPE are double trace operators, whose  OPE data admit an expansion similar to Eq.~\eqref{eq:c-expansion}, 
\begin{align}
\label{largeNexp}
\tau_{n, \ell}&=4+2n +\frac{1}{c} \gamma_{n,\ell}^{(1)}+\frac{1}{c^2} \gamma_{n,\ell}^{(2)}+\frac{1}{c^3} \gamma_{n,\ell}^{(3)}+ \dots \, ,\\ \label{largeNexp2}
a_{n,\ell}&=a_{n,\ell}^{(0)}+\frac{1}{c}a_{n,\ell}^{(1)}+\frac{1}{c^2}a_{n,\ell}^{(2)}+\frac{1}{c^3}a_{n,\ell}^{(3)}+\dots \, .
\end{align}
where with $\gamma_{n,\ell}^{(\kappa)}$ we will refer to the anomalous dimension at order $c^{-\kappa}$ and 
\begin{align}
\label{averageGamma}
\gamma_{n, \ell}^{(1)}&=-\frac{(n+1)(n+2)(n+3)(n+4)}{(\ell+1)(2n+\ell+6)}\, , \\
\label{averageA0}
a_{n,\ell}^{(0)}&=\frac{\pi (\ell+1)(\ell+2n+6)\Gamma(n+3)\Gamma(\ell+n+4)}{2^{(2\ell+4n+9)}\Gamma(n+\frac{5}{2})\Gamma(\ell+n+\frac{7}{2})}\, .
\end{align}
Generically double trace operators are constructed from combinations of the half-BPS operators in Eq.~\eqref{half-BPS},  they are schematically indicated as $[\mathcal{O}_p\, \mathcal{O}_p]_{n,\ell}=(\mathcal{O}_p \Box^{n}\partial_{\mu_1}\dots\partial_{\mu_\ell} \mathcal{O}_p -\text{traces})$ and  at leading order they  take their classical dimension $\Delta= 2p+2n+\ell$. From their definition, it is clear that starting from different $\mathcal{O}_p$ and with a different number of derivatives, one can construct more than one operator with the same twist and this will lead to a  degeneracy. More specifically, all operators of the form  $[\mathcal{O}_2\mathcal{O}_2 ]_{n,\ell}, [\mathcal{O}_3\mathcal{O}_3 ]_{n-1,\ell}, \dots, $ $[\mathcal{O}_{n+2}\mathcal{O}_{n+2} ]_{0,\ell}$  will contribute to Eq.~\eqref{largeNexp} and  having the same twist and spin and transforming under the same $SU(4)_R$ representation,  they give rise to a mixing problem. Let us denote the collection of  these $(n+1)$ degenerate states as $\mathcal{O}_{n, \ell, I}$ with $I=1, \dots, n+1$. Quite remarkably, this mixing has been partially solved up to order $c^{-1}$ \cite{Aprile:2017bgs,Aprile:2017xsp} and as a result $a_{n, \ell}^{(0)}$ and $\gamma_{n, \ell}^{(1)}$ are known for  each $I$,  concretely
\begin{align}
\gamma^{(1)}_{n, \ell, I}=\frac{-4(n+1)_4(\ell + 5 + 2 I)_{n+2-2I}}{(\ell - 1 + 2 I)_{n+2 - 2 I} (\ell + n + 1) (\ell +n + 6))} \, ,
\end{align}
where $(x)_n=\frac{(x+n-1)!}{(x-1)!}$ is the Pochhammer symbol. For the three-point function we have
\begin{align}
\braket{\mathcal{O}_2\mathcal{O}_2 \mathcal{O}_{n,\ell, I}}^2=a_{n,\ell, I}=a^{(0)}_{n, \ell} R_{n, \ell, I}c_{n, I} \, ,
\end{align}
where
\begin{align}\nonumber
R_{n, \ell, I}&=\frac{2^{-n-1} (4 I+2 \ell+3) (\ell+n+6)_{I-1}
   \left((I+\ell+1)_{-I+n+1}\right){}^{\text{sgn}(-I+n+1)}}{\left(I+\ell+\frac{5}{2}\right)_{n+
   1}}\, ,\\
   c_{n, I}&=\frac{2^{-n-1} \Gamma (2 I+3) \Gamma (n+1) \Gamma (-2 I+2 n+7)}{3 \Gamma (I) \Gamma
   (I+2) \Gamma (n+5) \Gamma (-I+n+2) \Gamma (-I+n+4)} \, .
\end{align}
In this context the expressions in Eq.~\eqref{averageGamma} and \eqref{averageA0} should be interpreted as averaged values over the index $I$.

Plugging in Eq.~\eqref{Hexpanion} the expansions \eqref{largeNexp}, \eqref{largeNexp2} one can obtain the expressions for $\mathcal{H}^{(\kappa)}(z, \zb)$ as infinite sums over the double trace spectrum, we report the explicit form for the  first few orders: 
\begin{align}
\mathcal{H}^{(0)}(z, \zb)&=(z \zb)^{n+2}a^{(0)}_{n, \ell} g_{2 n+8,\ell}(z, \zb)
\\
\mathcal{H}^{(1)}(z, \zb)&=(z \zb)^{n+2} \left(a^{(0)}_{n, \ell}\gamma^{(1)}_{n, \ell} \partial_n  +a^{(1)}_{n, \ell} \right)g_{2n+8,\ell}+\frac{1}{2} (z \zb)^{n+2} \log (z \zb)
   a^{(0)}_{n, \ell}\gamma^{(1)}_{n, \ell} g_{2n+8,\ell}
\\
\mathcal{H}^{(2)}(z, \zb)&=\frac{1}{2}(z \zb)^{n+2} \left(a^{(0)}_{n, \ell}  \left( \left(\gamma^{(1)}_{n, \ell}\right)^2 \partial^2_n  +2 \gamma^{(2)}_{n, \ell} \partial_n \right)+2 a^{(1)}_{n, \ell}\gamma^{(1)}_{n, \ell}
   \partial_n +2 a^{(2)}_{n, \ell} \right)g_{2n+8,\ell} \\ & \nonumber+\frac{1}{2} (z \zb)^{n+2} \log (z \zb) \left(\left(a^{(0)}_{n, \ell}
  \gamma^{(2)}_{n, \ell}+a^{(1)}_{n, \ell} \gamma^{(1)}_{n, \ell}\right)+a^{(0)}_{n, \ell}\left(\gamma^{(1)}_{n, \ell}\right)^2 \partial_n  \right)g_{2n+8,\ell}\\ & \nonumber +\frac{1}{8} (z \zb)^{n+2} \log ^2(z \zb) a^{(0)}_{n, \ell} \left(\gamma^{(1)}_{n, \ell}\right)^2g_{2n+8,\ell} 
\\
\mathcal{H}^{(3)}(z, \zb)&=\frac{1}{6} (z \zb)^{n+2} \left(6 a^{(0)}_{n, \ell}\gamma^{(1)}_{n, \ell}\gamma^{(2)}_{n, \ell} \partial^2_n 
  +a^{(0)}_{n, \ell} \left(\gamma^{(1)}_{n, \ell}\right)^3 \partial^3_n +6 a^{(0)}_{n, \ell} \gamma^{(3)}_{n, \ell} +\partial_n +\right. \\& \nonumber \left. \qquad 3
   a^{(1)}_{n, \ell}\left(\gamma^{(1)}_{n, \ell}\right)^2 \partial^2_n +6 a^{(1)}_{n, \ell}\gamma^{(2)}_{n, \ell}
   \partial_n +6 a^{(2)}_{n, \ell}\gamma^{(1)}_{n, \ell} \partial_n 
 +6 a^{(3)}_{n, \ell}\right) g_{2n+8,\ell}\\ & \nonumber +\frac{1}{4} (z \zb)^{n+2} \log
   (z \zb) \left(2\left(a^{(0)}_{n, \ell}\gamma^{(3)}_{n, \ell}+a^{(1)}_{n, \ell}\gamma^{(2)}_{n, \ell}+a^{(2)}_{n, \ell}\gamma^{(1)}_{n, \ell}\right)+\right. \\ & \left. \nonumber \qquad \gamma^{(1)}_{n, \ell} \left(a^{(0)}_{n, \ell}\left(\gamma^{(1)}_{n, \ell}\right)^2 \left(\partial^2_n +4\gamma^{(2)}_{n, \ell}
   \partial_n \right)+2 a^{(1)}_{n, \ell}\gamma^{(1)}_{n, \ell} \partial_n\right)\right)g_{2n+8,\ell}  \\ & \nonumber +\frac{1}{8} (z \zb)^{n+2} \log ^2(z \zb)
  \gamma^{(1)}_{n, \ell} \left( \left(2 a^{(0)}_{n, \ell} \gamma^{(2)}_{n, \ell}+a^{(1)}_{n, \ell}\gamma^{(1)}_{n, \ell}\right)+a^{(0)}_{n, \ell} \left(\gamma^{(1)}_{n, \ell}\right)^2\partial_n\right)g_{2n+8,\ell}\\ & \nonumber +\frac{1}{48} (z \zb)^{n+2} \log ^3(z \zb)
   a^{(0)}_{n, \ell} \left(\gamma^{(1)}_{n, \ell}\right)^3 g_{2n+8,\ell}
\end{align}

The knowledge of the leading corrections to the dimension and OPE coefficient of the degenerate states has been enough to give the full four-point function at order $c^{-2}$ \cite{Alday:2017vkk,Alday:2017xua, Aprile:2017bgs}, generically this does not hold at any given order in the perturbative expansion. However, a closer look to the expressions above reveals that only tree-level data are needed to compute what we called the {\textit{leading logarithmic singularities}} at each $\mathcal{H}^{(\kappa)}$.  Concretely, they take the general form
\begin{align}
\label{eq:leading logs}
\mathcal{H}^{(\kappa)}(z,\zb)\Big |_{\log ^{\kappa}(z \zb)}=\frac{1}{2^{\kappa} \kappa!}\sum_{n,\ell} \sum_{I=1}^{n+1} (z \zb)^{n+2} a^{(0)}_{n,\ell, I}(\gamma_{n,\ell, I}^{(1)})^{\kappa}g_{2n+8, \ell}(z, \zb) \, ,
\end{align}
where only tree-level data of double trace operators enter and where we have reintroduced the explicit dependence on the mixing index $I$.

\subsection{Leading logarithmic singularities}
In principle we have all the information to compute the leading logarithmic terms in $\braket{\mathcal{O}_2\mathcal{O}_2\mathcal{O}_2\mathcal{O}_2}$. The difficulty is performing the sums over $n$, $\ell$ and $I$. To do so we will use the method introduced in \cite{Caron-Huot:2018kta}. Before reviewing it, let us first comment on the general analytic structure of the correlator.  Holographic CFTs with a bulk gravity dual are expected to develop a singularity, the bulk point singularity, which is probed by sending $z \to \zb$ \cite{Alday:2018pdi, Heemskerk:2009pn,Okuda:2010ym, Maldacena:2015iua, Gary:2009ae}, and whose behaviour is determined from the large $n$ limit of $\gamma_{n, \ell}$.  For the terms that we are going to study, one finds
\begin{align}
\mathcal{H}^{(\kappa)}\Big|_{\log^\kappa z \zb}=\frac{(z \zb)^2}{(z-\zb)^{a+4}} f^{(\kappa)}(z, \zb)
\end{align}
where $f^{(\kappa)}$ contains functions of maximal degree of transcendentality $\kappa$ with polynomial coefficients and the power $a$ depends on the order of expansion as
\begin{align}
\label{alpha}
a=3\underbrace{\kappa}_{\text{\# of AdS loops}}+5\underbrace{(\kappa-1)}_{\text{\# of S loops}} \, .
\end{align}
Since from Eq.~\eqref{averageGamma} $\gamma^{(1)}_{n, \ell} \sim n^3$ for $n \gg 1$, one would have naively expected only the $3 \kappa$ term, where $\kappa$ parametrizing the order in the $c$ expansion can be equally viewed as counting the number of loops in AdS. However, the degeneracy among the double trace operators and the consequent presence of mixing  enhance this singular behaviour determining the additional factor in Eq.~\eqref{alpha}, which accounts for the degrees of freedom of the internal S$^5$ manifold. 

Let us now turn to a brief description of the results in  \cite{Caron-Huot:2018kta} and let us start by introducing an eight-order differential operator $\Delta^{(8)}$\footnote{With respect to Eq.~(3.15) in \cite{Caron-Huot:2018kta} we are restricting directly to the case $r=0=s$ and no dependence on the $SU(4)_R$ cross-ratios. This simplification applies to the case of a correlator of four identical $\mathcal{O}_2$ operators.}:
\begin{align}
\label{delta8}
\Delta^{(8)} \equiv \frac{z \zb}{(z-\zb)}\mathcal{D}_z(\mathcal{D}_z-2)\mathcal{D}_{\zb} (\mathcal{D}_{\zb}-2)\frac{(z-\zb)}{z \zb} \, ,
\end{align}
where 
\begin{align}
\mathcal{D}_x\equiv x^2 \partial_x (1-x)\partial_x \, .
\end{align}
The importance of this operator relies on the fact that when it acts on superconformal blocks, as shown in \cite{Aprile:2018efk, Caron-Huot:2018kta}, its eigenvalues are equal to half of the anomalous dimension of double trace operators, i.e.\footnote{It is convenient to rewrite \begin{align}
\gamma_{n, \ell, I}^{(1)}=\frac{-4(n+1)_4(\ell+n+2)_4}{(\ell+2I-1)_6} \, .
\end{align}}
\begin{align}
\Delta^{(8)}\left(u^{\tau/2}g_{\tau+4,\ell}(z, \zb) \right)=-2(n+1)_4(\ell+n+2)_4 \left(u^{\tau/2}g_{\tau+4,\ell}(z, \zb) \right) \, .
\end{align}
This remarkable property is conjectured to descend from an accidental 10-dimensional conformal symmetry $SO(10,2)$  of the underlying supergravity amplitudes. Motivated by this symmetry, one discovers that ten dimensional blocks actually diagonalize the mixing problem at order $c^{-1}$ and can be used to express the leading logarithmic terms in Eq.~\eqref{eq:leading logs} in a closed form, see \cite{Caron-Huot:2018kta} for more details,
\begin{align}
\label{Hwithh}
\mathcal{H}^{(\kappa)}(z,\zb)\Big |_{\log^{\kappa}(z \zb)}=\left[\Delta^{(8)} \right]^{\kappa-1}\left( \mathcal{D}_{(3)} h^{(\kappa)}(z)+ z \leftrightarrow \zb \right) \, ,
\end{align}
where the operator $\mathcal{D}_{(3)}$ is used to build ten-dimensional blocks and it is defined as
\begin{align}
\label{eq:D3Op}
\mathcal{D}_{(3)}=\left(\frac{z \zb}{\zb-z}\right)^7+\left(\frac{z \zb}{\zb-z}\right)^6\frac{z^2}{2}\partial_z+\left(\frac{z \zb}{\zb-z}\right)^5\frac{z^3}{10}\partial_z^2 z+\left(\frac{z \zb}{\zb-z}\right)^4\frac{z^4}{120}\partial_z^3 z^2 \, .
\end{align}
Finally,
\begin{align}
\label{hform}
h^{(\kappa)}(z)=\frac{1}{\kappa!}\left(-\frac{1}{2}\right)^\kappa \sum _{\ell=0, \text{ even}}^{\infty } \frac{960 \Gamma(\ell+1)\Gamma(\ell+4) z^{\ell+1}  \, _2F_1(\ell+1,\ell+4;2 \ell+8;z)}{\Gamma (2 \ell+7)\left[(\ell+1)_6\right]^{\kappa-1}} \, .
\end{align}
Thus in principle we have an explicit expression to compute higher logarithmic terms at a given loop order. Although we were able to develop a simple algorithm to find all the $h^{(\kappa)}$ resumming Eq.~\eqref{hform},  applying $\Delta^{(8)}$ still remains a technically involving task, given the growing number of derivatives one needs to take. However, as we will discuss in the following, no derivatives are needed once one is only interested  in the bulk point limit of the correlation function. 
\subsection{Double Discontinuity and flat space limit}
\label{subsec:DDandFlat}
The study of the flat space or equivalently the bulk point limit of correlation functions has been initiated in the works  \cite{Gary:2009ae, Heemskerk:2009pn,Maldacena:2015iua} and it is interesting for at least two reasons. First of all  it can provide an insight on the analytic structure of the four-point function, secondly in this regime and for infinite 't Hooft coupling, CFT correlators can be mapped to scattering amplitudes in type IIB supergravity in the flat $\mathbb{R}^{10}$, which are in general more accessible and easier to compute. Our goal is to provide a general formula for the leading logarithmic term in this limit and at any order in the $c^{-\kappa}$ expansion. In doing so, we will generalize the results in \cite{Alday:2017vkk}, where it is shown that at one loop, there exists a one to one correspondence between the \textit{discontinuity} of the non-analytic piece of the four-graviton amplitude in $\mathbb{R}^{10}$ and the flat space limit of the \textit{double discontinuity} (dDisc) of $\mathcal{H}^{(2)}(z, \zb)$. 

{\bf{Double Discontinuity}}\\
The dDisc of an amplitude $\mathcal{H}(z, \zb)$ is defined as the difference between the Euclidean correlator and its two possible analytic continuations around $\zb=1$, keeping $z$ fixed:
\begin{align}
\label{dDiscLog}
\text{dDisc} \mathcal{H}(z, \zb) \equiv \mathcal{H}(z, \zb) -\frac{1}{2}\left( \mathcal{H}^{\circlearrowleft}(z,\zb)+\mathcal{H}^{\circlearrowright}(z, \zb) \right) \, .
\end{align}
This quantity can be computed exploiting crossing symmetry and passing to the $t$-channel, where it acts trivially. Here, the only terms in the correlator $\mathcal{H}^{(k)}$ with non vanishing dDisc are $\log^n(1-\zb)$, with $n \geq 2$ ($n\leq \kappa$), which are completely determined by OPE data at order $\kappa-1$. Then, through the Lorentzian inversion formula \cite{Caron-Huot:2017vep}, this quantity allows to reconstruct the full correlation function.  By crossing $\log(1-\zb)$ corresponds to $\log (z \zb)$, hence at each order the leading logarithmic term in Eq.~\eqref{eq:leading logs} will contribute to dDisc as
\begin{align} 
\text{dDisc} [\log^{\kappa}(1-\zb)]&= 2 \pi^2   \kappa(\kappa -1) \log^{\kappa-2} (1-\zb) +\text{ lower powers of }\log(1-\zb) \, .
\end{align}
Unfortunately for $\kappa \geq 3$, also lower transcendental pieces contributes to the dDisc and other contributions should be considered to reconstruct the full correlator, even in the flat space limit. For $\kappa \geq 3$ an additional obstacle to the complete reconstruction of $\mathcal{H}^{(k)}$ is the appearance of multi-trace operators, besides the double trace that we are already considering. Their emergence should become clearer later on in our discussion, when we will study the dual scattering amplitude at two loops and we will analyse the origin of its discontinuities. The $\kappa=2$ example in \cite{Alday:2017vkk, Alday:2018pdi} is eventually a very special one, in this case in fact the only term entering dDisc is \eqref{eq:leading logs}, thus allowing to successfully compute the whole $\mathcal{H}^{(2)}$. The main focus of this paper is to understand what can be inferred from the knowledge of this piece of information at higher order and which conclusions can we draw about its relation to the four-graviton amplitude. 

{\bf{Bulk point limit}}\\
Let us now pass to properly discuss the \textit{flat space} or \textit{bulk point limit} of  correlators and its implications both on the CFT and on the gravity side. Consider sufficiently localized wave-packets in AdS, such that it can be approximated with flat space, then bulk S-matrix elements in this limit can be reproduced by correlators in the boundary CFT with a prescribed singular behaviour. In particular, to reproduce the bulk kinematics, the CFT correlation function must diverge at $z \to \zb$. Notice that this type of singularity can manifest itself only in  Lorentzian signature, where $z$ and $\zb$ are real and independent variables, so to reach the bulk point limit it is needed to  analytically continue the Euclidean correlator to imaginary times and then take this Lorentzian time to $\pi$. In doing so,  the correlation function picks an additional phase, while the OPE expansion still holds,
\begin{align}
\mathcal{H}(z, \zb)_{\text{Lorentzian}}=\sum_{\Delta, \ell} e^{-i \pi (\Delta-\ell)}a_{\Delta, \ell} g_{\Delta, \ell}(z, \zb) \, .
\end{align}
In large $N$ theories, where this sum is over an infinite tower of double trace operators, the phase is responsible for the appearance of a singularity, which turns out to be dominated by the large-$n$ tail of the sum \cite{Alday:2017vkk}. In formulae \cite{Heemskerk:2009pn}:
\begin{align}
\label{correlatorExpansion}
z \zb (\zb-z)\, \mathcal{H}_{\text{Lorentzian}}(z, \zb) \approx -64 i \pi \sum_{n} n^2 e^{2 i x n} \sum_{\ell\text{ even}} (\ell+1)^2 P_{\ell}(\cos\theta) \frac{\braket{a e^{-i \pi \gamma}}_{n, \ell}}{\braket{a^{(0)}}_{n, \ell}} \, ,
\end{align}
where $x \sim z - \zb \to 0$ parametrizes how to approach the singularity  and $P_{\ell}$ are four dimensional Legendre polynomials. The left hand side of Eq.~\eqref{correlatorExpansion} is the quantity that we are going to relate to the scattering amplitude in the dual gravity theory. In the right hand side instead, we recognize  the familiar partial wave expansion valid for amplitudes in generic quantum field theory, thus motivating  the following relation: 
\begin{align}
\label{eq:CFTGravityflat}
\lim_{n\to\infty}\frac{\braket{a e^{- i \pi \gamma}}_{n,\ell}}{\braket{a^{(0)}}_{n,\ell}}=b_\ell(s), \qquad L \sqrt{s}= 2 n
\end{align}
where $b_\ell(s)$ are the coefficients of the partial wave expansion of the flat space gravity amplitude, we will come back to them in Sec.~\ref{sec:Amplitude}.\\
Let us summarize: so far we have seen that to determine the singular behaviour of the CFT correlator in the bulk point regime we need to study averages (over mixing index) of the type $\braket{a e^{-i \pi \gamma}}$ in the large $n$ limit. On the gravity side for weakly coupled theories, these singularities correspond to \textit{bulk Landau diagrams} (see \cite{Maldacena:2015iua} for details) which can be computed as in flat space time. Finally, the linking point between CFT and gravity is given by Eq.~\eqref{eq:CFTGravityflat}.

Given (\ref{largeNexp}) and (\ref{largeNexp2}), the average $\braket{ae^{-i \pi \gamma}}$ admits the large $c$ expansion
\begin{align} \label{expagamma} \nonumber
\braket{ae^{-i \pi \gamma}}_{n,\ell}&=\braket{a^{(0)}}+c^{-1}\left( \braket{a^{(1)}}-i \pi\braket{a^{(0)}\gamma^{(1)}} \right)+  c^{-2}\left(\braket{a^{(2)}}-i \pi\braket{a^{(1)}\gamma^{(1)}+a^{(0)}\gamma^{(2)}} \right.\\ & \left. \nonumber -\frac{\pi^2}{2}\braket{a^{(0)}{\gamma^{(1)}}^2}\right) +c^{-3}\left(\braket{a^{(3)}}
 -i \pi\braket{a^{(2)}\gamma^{(1)}+a^{(1)}\gamma^{(2)}+a^{(0)}\gamma^{(3)}}\right.\\ & \left. -\pi^2\braket{\frac{a^{(1)}{\gamma^{(1)}}^2}{2}+a^{(0)}\gamma^{(1)}\gamma^{(2)}} +\frac{i \pi^3}{6}\braket{a^{(0)}{\gamma^{(1)}}^3} \right)+\mathcal{O}(c^{-4})\, .
\end{align}
The large $n$ limit can then be taken using directly the inversion formula  and results in an expression depending solely on the dDisc defined above \cite{Alday:2017vkk},
\begin{align} \label{agamma}
 &\nonumber \frac{\braket{a e^{-i \pi \gamma}}_{n, \ell}}{ \braket{a^{(0)}}_{n, \ell}} \xrightarrow{n\gg1}1+ \frac{i \pi n^{3}c^{-1}}{2(l+1)}+\sum_{\kappa=2} \frac{c^{-\kappa}}{n^2 (\ell+1)} \int_C \frac{dx}{2 \pi i}e^{-2n x}\\&  \qquad \qquad \quad \times \int_0^1\frac{d\zb}{\zb^2}\left(\frac{1-\sqrt{1-\zb}}{1+\sqrt{1-\zb}}\right)^{l+1}\frac{\text{dDisc}\left[z \zb (\zb-z) \mathcal{H}^{(\kappa)}( z^\circlearrowright,\zb)\right]}{4 \pi^2} \, .
\end{align}
Finally the flat space limit is realised by taking $z=\zb+2x \zb \sqrt{1-\zb}$ with $x \rightarrow 0$ after having analytically continued $z$ around $0$ ($z ^\circlearrowright$ in the formula above stands exactly for the continuation  $z\rightarrow z\, e^{-2 \pi i }$). The $x$-integral has the effect of converting powers of  $\frac{1}{z-\zb}$, which are related to the expansion order $\kappa$ through \eqref{alpha}, into powers of $n$, and consequently of $L\sqrt{s}$ according to Eq.~\eqref{eq:CFTGravityflat}.  Finally in this integral above $C$ is a key-hole contour encircling the origin clockwise.
\subsection{Flat space limit of higher logarithmic terms at all loops}\label{sec_alllcft}
\label{sec:FSLeadingLogs}
The relation \eqref{agamma} assumes the knowledge of the full double discontinuity of the correlator, however, as anticipated before, Eq.~\eqref{Hwithh} is sufficient to only partially fix it for $\kappa > 2$, thus making impossible to completely use  this expression and a direct comparison with the amplitude. However as we have seen explicitly for the $\kappa=3$ case in \cite{Bissi:2020wtv}, which is reviewed in Appendix \ref{sec:2loopEx}, something interesting can still be said also for these higher logarithmic terms.

Let us start by taking a closer look at the expression in \eqref{Hwithh}: in the $x \to 0$ limit the dominant terms are the most divergent ones as $z\to \zb$  (i.e.  the highest powers of $\frac{1}{z-\zb}$).  This observation greatly simplifies our computation,  especially  when applying $\mathcal{D}_{(3)}$ in Eq.~\eqref{eq:D3Op} and $\Delta^{(8)}$ in Eq.~\eqref{delta8}. Indeed the evaluation of Eq.~\eqref{Hwithh} reduces to:
\begin{align}
&\left\lbrace\mathcal{D}_{(3)}h^{(\kappa)}(z)+(z \leftrightarrow \zb)\right\rbrace \xrightarrow{\text{flat space}} \left(\frac{z \zb}{z-\zb}\right)^7\left(h^{(\kappa)}(z)-h^{(\kappa)}(\zb)\right)\, ,\\
\label{eq:Delta8flat}
&\left[\Delta^{(8)}\right]^{\kappa-1}\left\lbrace\mathcal{D}_{(3)} h^{(\kappa)}(z)+(z \leftrightarrow \zb)\right\rbrace  \xrightarrow{\text{f.s.}}  \frac{7\cdot 3^{2 \kappa-1} 64^{\kappa-1}(z \zb)^7}{(z-\zb)^{8\kappa-1}} w_\kappa(z,\zb)\left( h^{(\kappa)}(z)-h^{(\kappa)}(\zb) \right) \, .
\end{align}
where in the second line we have divided everything by $(z \zb)^2$ to adjust it to our normalization. Notice that the  power of $(z-\zb)$ correctly reproduces our prediction in   Eq.~\eqref{alpha}  ($8\kappa-1=a+4$). $w_\kappa(z, \zb)$ is a polynomial of degree $4 (3 \kappa-4)$, which, in the limit $z=\zb$ we are working in, reduces to
\begin{align}
\label{fkappa}
w_\kappa(z, \zb) \xrightarrow{z=\zb}\frac{1}{35} 8^{1-2 \kappa} 9^{-\kappa} \Gamma (8 \kappa-2) (1-\zb)^{4(\kappa-1)} \zb^{4 (2\kappa-3)}\, .
\end{align}
Putting everything together, one finally finds
\begin{align}
\label{FlatSpaceh}
&\frac{\text{dDisc}[z \zb (\zb-z)]\mathcal{H}^{(\kappa)}|_{\log^{\kappa}(z, \zb)}(z^{\circlearrowright}, \zb)]}{4\pi^2} \xrightarrow{\text{flast space}} \frac{\Gamma[8\kappa-2]}{(2x)^{8\kappa-2}}\frac{(1-\zb)^{4\kappa+3}}{120 \zb^{4\kappa}} \times \\ \nonumber
& \qquad \times \frac{1}{4 \pi^2}\text{dDisc}\left[\frac{(z \zb)^2}{(1-z)^2(1-\zb)^2}\log^\kappa((1-z)(1- \zb))\left(h^{(\kappa)}(1-z^{\circlearrowright})-h^{(\kappa)}(1-\zb) \right) \right]_{z=\zb}\\ \nonumber  & \qquad \equiv \frac{\text{dDisc}\left[\log^{\kappa}((1-\zb)(1-z)\right]_{z=\zb}}{4\pi^2} g^{(\kappa)}(\zb) \, ,
\end{align}
where we have explicitly used crossing symmetry to pass to the $t$-channel. Notice that since one has to take $z=\zb$, the only terms surviving in the difference are the ones coming from the analytic continuation of $z$ around the origin. Moreover, given that only integer powers of $(1-\zb)$ appear and $h^{(\kappa)}$ can not contain $\log(1-\zb)$, dDisc acts non trivially only on $\log^{\kappa}(1-\zb)$, thus returning  Eq.~\eqref{dDiscLog}. The last point that we want to stress is that, as it is evident from \eqref{FlatSpaceh}, in the flat space limit we do not have to take any derivatives, streamlining a lot the computations and eventually the only information one needs to know is an explicit expression for the $h^{(\kappa)}$'s. Eq.~\eqref{hform} gives a very compact prescription for finding them, but it involves an infinite sum, which is not straightforward to solve. Quite remarkably, we have found a simple algorithm to compute this quantity, allowing us to study different cases\footnote{We have computed the full form of $h^{(\kappa)}$ up to $\kappa=7$, their expressions can be found in the ancillary Mathematica file, and for the highest transcendental pieces up to $\kappa=21$.} and deduce a general form at any order in $c$. 

First of all we have noticed the general structure
 \begin{align}
\label{eq:hsmall}
&h^{(\kappa)}(z)=j_0(z)+j_1(z)H_{1}(z)+\left(j_2(z)K_2(z)+j_2\left(\frac{z}{z-1}\right)I_2(z)\right)\\ & \nonumber  +\left(j_3\left(\frac{z}{z-1}\right)K_3(z)+j_3(z)I_3(z)\right)+ \cdots+ \begin{cases}
\left(j_\kappa(z)K_\kappa(z)+j_\kappa\left(\frac{z}{z-1}\right)I_\kappa(z)\right) & \text{for } \kappa\text{ even}\\
 \left(j_\kappa\left(\frac{z}{z-1}\right)K_{\kappa}(z)+j_\kappa(z)I_\kappa(z)\right)& \text{for } \kappa\text{ odd}
\end{cases}
\end{align}
where $j_i$ are generic function, $I_n$ and $K_n$ are iterative integrals defined as
\begin{align}
&I_n(z)=\int_{0}^{z}\frac{dz^{\prime}}{z^{\prime}(1-z^{\prime})^{(n-1)\text{mod}_2}}I_{n-1}(z^{\prime}) \hspace{1.3cm} \text{where} \qquad \quad I_1(z)\equiv H_1(z) \, ,\\
&K_n(z)=\int_{0}^{z}\frac{dz^{\prime}}{z^{\prime}(1-z^{\prime})^{n\text{mod}_2}}K_{n-1}(z^{\prime})  \hspace{1.6cm} \text{where} \qquad \quad K_1(z)\equiv H_1(z) \, .
\end{align}
Here and in the following we will denote with $H_{\dots}(x)$ the Harmonic Polylogarithms (HPL), which are defined in Appendix \ref{sec:HarmonicPolylog}. It follows directly from their definition that, both for the $I$ and $K$ integral, at each step $n$ we are adding alternatively  a $0$ or  a $0$ and a $1$ to each HPL appearing at order $n-1$, i.e. $H_{a_1 \dots a_{n-1}}(z) \to H_{0a_1 \dots a_{n-1}}(z)$ or $H_{a_1 \dots a_{n-1}}(z) \to H_{0a_1 \dots a_{n-1}}(z)+H_{1a_1 \dots a_{n-1}}(z)$. Hence, starting from $H_{1}(z)$, a one is always followed by a zero and there are no two consecutive ones to the left of the index vector; the only exception  is $\HPL{11}{z}$ in $I_2$. This is enough to state that the maximal logarithmic divergence\footnote{Remember that $h^{(k)}$ does not contain $\log z$ because by definition they are already factored out in the correlation function.} of $\sim \log (1-z)$-type that $h^{(\kappa)}$ can develop is precisely $ \HPL{11}{z}$ and this term produces a $\log^2 V$ in the small $V$ expansion of the correlator. The full leading logarithmic term will eventually inherit the same behaviour, even away from the flat space limit, so we expect 
\begin{align} \label{leadLogUV}
\text{leading logs in }\mathcal{H}^{(\kappa)} \sim \log^{\kappa} U \log^2 V \, .
\end{align}
In this way we have explicitly checked to all order the conjectures in \cite{Bissi:2020wtv}, which was originally inspired by  a diagrammatic interpretation in the dual gravity theory, see also \cite{Liu:2018jhs}.\\
To conclude, from the study of  different explicit examples we are able to find a close expression for the  two highest transcendentality terms in~\eqref{eq:hsmall},
\begin{align}
\label{pandqPol}
\begin{cases}
p_{\kappa}(z)K_{\kappa}(z)+p_\kappa\left(\frac{z}{z-1}\right) I_{\kappa}(z)+q_\kappa\left(\frac{z}{z-1}\right)K_{\kappa-1}(z)+q_\kappa(z) I_{\kappa-1}(z) & \qquad \kappa\text{ even}  \\
p_\kappa\left(\frac{z}{z-1}\right)K_\kappa(z)+p_\kappa(z) I_\kappa(z)+q_{\kappa}(z)K_{\kappa-1}(z)+q_\kappa\left(\frac{z}{z-1}\right) I_{\kappa-1}(z) & \qquad \kappa\text{ odd} \\
\end{cases}
\end{align}
where the polynomials are defined as
\begin{align}
\label{pPol}
p_{\kappa}(z)&=\frac{(-1)^{\kappa } 4^{3-2 \kappa } 15^{1-\kappa }}{z^5 \kappa !} \left(\left(-6\ 5^{\kappa }+10^{\kappa }+50\right)
   z^2+3 \left(5^{\kappa }-25\right) z+30\right) \, ,\\ \label{qPol}
\nonumber    q_{\kappa}(z)&=\frac{\left(-\frac{1}{3}\right)^{\kappa } 2^{4-4 \kappa } 5^{1-\kappa }}{z^5 \kappa !} \left \lbrace -12 \left(-3\ 5^{\kappa }+10^{\kappa }+5\right) z^4+\left(-54\ 5^{\kappa }+3\ 10^{\kappa
   +1}-30\right) z^3 \right. \\&  \left. \qquad +\left(6 \left(17\ 5^{\kappa }+10^{\kappa }+265\right)-2 \left(39\ 5^{\kappa
   }+2^{\kappa +1} 5^{\kappa }+685\right) \kappa \right) z^2 \right. \\& \nonumber \left. \qquad+\left(3 \left(13\ 5^{\kappa }+685\right)
   \kappa -60 \left(5^{\kappa }+43\right)\right) z+1104  -822 \kappa\right\rbrace \, .
\end{align}
In Appendix~\ref{appendix:polStructure} we find the same polynomials starting this time from the amplitude and we will explain how to connect the results from both sides\footnote{For the highest transcendental term an expression was found also in \cite{Aprile:2018efk} and it matches the one we obtained.}.

\section{Four-graviton scattering amplitude} \label{sec:Amplitude}
\begin{figure}
\centering 
\subfloat[][]{
\centering
\begin{tikzpicture}[scale=0.4]
       \draw [thick] (0,0) -- (3,0)--(3,3)--(0,3)--(0,0);
    \draw [thick] (0,0) -- (-1,-1);
    \draw [thick] (0,3) -- (-1,4);
    \draw [thick] (3,0) -- (4,-1);
    \draw [thick] (3,3) -- (4,4);
    \draw [thick,->] (3,0) -- (1.5,0);
    \node [below left=0.1 cm] at (-1,-1)  {$p_1$};
    \node [below right=0.1 cm] at (4,-1)   {$p_4$};
    \node [above right=0.1cm] at (4, 4) {$p_3$};
    \node [above left=0.1 cm] at (-1,4)  {$p_2$};
    \node [below=0.1 cm] at (1.5,0)  {$l_1$};
\end{tikzpicture}}
\qquad
\subfloat[][]{
\centering
\begin{tikzpicture}[scale=0.4]
    \draw [thick] (0,0) -- (6,0)--(6,3)--(0,3)--(0,0);
    \draw [thick] (3,0) -- (3,3);
    \draw [thick] (0,0) -- (-1,-1);
    \draw [thick] (0,3) -- (-1,4);
    \draw [thick] (6,0) -- (7,-1);
    \draw [thick] (6,3) -- (7,4);
    \draw [thick,->] (3,0) -- (4.5,0);
    \draw [thick,->] (3,0) -- (1.5,0);
    \node [below left=0.1 cm] at (-1,-1)  {$p_1$};
    \node [below right=0.1 cm] at (7,-1)   {$p_4$};
    \node [above right=0.1cm] at (7, 4) {$p_3$};
    \node [above left=0.1 cm] at (-1,4)  {$p_2$};
    \node [below=0.1 cm] at (1.5,0)  {$l_1$};
    \node [below=0.1 cm] at (4.5,0)  {$l_2$};
\end{tikzpicture}}
\qquad
\subfloat[][]{
\centering
\begin{tikzpicture}[scale=0.4]
    \draw [thick] (6,0) -- (0,0)--(0,3)--(6,3);
    \draw [thick] (3,0) -- (6,3);
    \draw [thick] (3,3) -- (4.3,1.7);
    \draw [thick] (4.7,1.3) -- (6,0);
    \draw [thick] (0,0) -- (-1,-1);
    \draw [thick] (0,3) -- (-1,4);
    \draw [thick] (6,0) -- (7,-1);
    \draw [thick] (6,3) -- (7,4);
    \draw [thick,->] (3,0) -- (4.5,0);
    \draw [thick,->] (3,0) -- (1.5,0);
    \node [below left=0.1 cm] at (-1,-1)  {$p_1$};
    \node [below right=0.1 cm] at (7,-1)   {$p_4$};
    \node [above right=0.1cm] at (7, 4) {$p_3$};
    \node [above left=0.1 cm] at (-1,4)  {$p_2$};
    \node [below=0.1 cm] at (1.5,0)  {$l_1$};
    \node [below=0.1 cm] at (4.5,0)  {$l_2$};
\end{tikzpicture}}
\caption{One loop planar box digram (a) and two loop planar (a) and non planar (b) double box diagram.}
\label{Fig:boxes}
\end{figure}
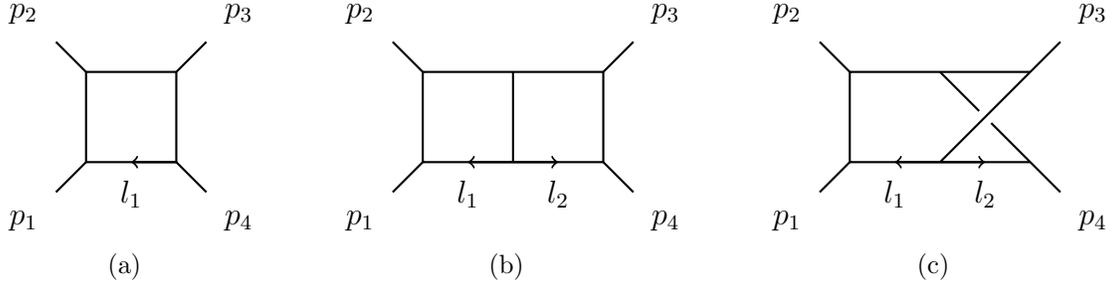

It is well known that $\frac{1}{N}$ corrections to the four point functions of four $\mathcal{O}_2$ half-BPS operators in $\mathcal{N}=4$ SYM in the $\lambda \to \infty$ limit are dual to perturbative scattering amplitudes of gravitons in IIB supergravity on AdS$_5 \times$S$^5$. For the purposes of studying the CFT bulk-point limit, the background geometry can be effectively taken to be ten dimensional flat space, so that the final object we are going to study is $\mathcal{A}_{10}^{sugra}$,  denoting the supergravity amplitude of four gravitons in $\mathbb{R}^{10}$. Similarly to the $1/c$ expansion in Eq.~\eqref{eq:c-expansion}, this quantity admits a loop expansion in powers of the gravitational constant $G_N$ and up to two loops this reads
\begin{align}
\label{10dAmpl}
\mathcal{A}_{10}^{sugra}&=\hat{K} \left\lbrace\frac{8 \pi G_N}{s t u}+\left(8 \pi G_N\right)^2\left(I_{box}(s,t)+I_{box}(s,u)+I_{box}(t,u) \right)\right. \\& \left.\nonumber + \left(8 \pi G_N\right)^3\left(s^2\left(I_{db}^{pl}(s,t)+I_{db}^{np}(s,t)+t\leftrightarrow u\right)+t^2\left(I_{db}^{pl}(t,s)+I_{db}^{np}(t,s)+s\leftrightarrow u\right)
\right.\right. \\& \left.\left.\nonumber+u^2\left(I_{db}^{pl}(u,s)+I_{db}^{np}(u,s)+s\leftrightarrow t\right) \right)+ \mathcal{O}(G_N^4 ) \right\rbrace\\
\label{amplitudeinL}
&\quad \equiv (\pi L)^5 s^4\Big\lbrace \frac{L^3 F_1(x)}{s^3 c}+\frac{L^{11} s F_2(x)}{c^2}+\frac{L^{19} s^5 F_3(x)}{c^3}+\mathcal{O}(c^{-4}) \Big\rbrace \, ,
\end{align}
where we have reported the value of the amplitude at tree-level, while at higher loops we have substituted it with the sum of the corresponding Feynman diagrams. Here $I_{box}$, $I_{db}^{pl}$ and $I_{db}^{np}$ indicate respectively the one-loop box diagram, planar and non planar double box  in Fig.~\ref{Fig:boxes}. With $s, t$ and $u$ we denote the usual Mandelstam variables satisfying the constraint $s+t+u=0$. $\hat{K}$ is a dimension-eight kinematic factor, depending on the  polarizations of the gravitons. In order to compare the amplitude with $\mathcal{G}(U, V)$, these polarization vectors should be chosen  to be null tensors and orthogonal to all the external momenta; in this configuration $\hat{K}$ reduces to  an overall  $s^4$ factor. In the second line we have used the identification $8\pi G_N=\pi^5L^8c^{-1}$ to make manifest the relation with the large $c$ expansion with $L=L_{\text{AdS}_5}=L_{\text{S}^5}$  the AdS and sphere radii. The $F_{\kappa}$ are functions of the adimensional parameter $-\frac{t}{s}=1-x =\frac{1-\cos\theta}{2}$, which has been introduced to make the  $s$ and $L$ dependence manifest. These factors are in fact the ones related to the large $n$ behaviour of the correlator, as prescribed by Eq.~\eqref{eq:CFTGravityflat}.

The five dimensional $\mathcal{A}_5$ amplitude is obtained by dividing $\mathcal{A}^{sugra}_{10}$ by the sphere volume $\pi^3 L^5$ and it can be expanded in partial waves as
\begin{align}
\label{amplitudeExp}
i \mathcal{A}_5(s,t)=\frac{128\pi}{\sqrt{s}} \sum_{\ell \text{ even}} (\ell +1)^2 b_{\ell}(s)P_{\ell}(\cos\theta) \, .
\end{align}
Using the orthonormality of the Legendre polynomials, one can extract the coefficients as
\begin{align}
\label{bell}
b_\ell(s)&=1+\frac{i \pi}{\ell+1}\int_0^{\pi}\frac{d\theta}{\pi}\sin\theta\sin((\ell+1)\theta)\frac{\sqrt{s}\mathcal{A}_5(s, \cos\theta)}{64\pi^2}\\ \nonumber
&=1+ \left(\frac{L\sqrt{s}}{2}\right)^3\frac{i \pi c^{-1}}{2(\ell+1)}+\sum_{\kappa=2}\frac{2 \pi i c^{-\kappa}}{\ell+1}\int_0^1\frac{d\zb}{\zb^2}\left(\frac{1-\sqrt{1-\zb}}{1+\sqrt{1-\zb}}\right)^{\ell+1}  \\ & \qquad \qquad  \qquad  \qquad \nonumber  \times \left(\text{disc}_t+(-1)^\ell\text{disc}_u \right)\frac{\sqrt{s}\mathcal{A}_5^{(\kappa)}(s, \cos\theta)}{64\pi^2} \, ,
\end{align} 
where we have defined $\mathcal{A}_5^{(\kappa)}$ as the $(\kappa-1)$-loop amplitude.
In the second line we have plugged in the explicit expression for $\mathcal{A}_5^{tree}$, while for the other orders  we have expressed the amplitude through its discontinuities  via a dispersion relation \cite{Caron-Huot:2017vep,Alday:2017vkk}. In the cases at hand, $\ell$ runs over even integers, so that the discontinuity in the $t$-channel (disc$_t$) and the one in the $u$-channel (disc$_u$), taken at fixed $s$, actually sum up. \\
The expressions in Eq.~\eqref{amplitudeExp} and \eqref{bell} are very reminiscent of Eq.~\eqref{correlatorExpansion} and \eqref{agamma}, thus motivating the following identifications:
\begin{gather}\label{identification_cft_sugra1}
\frac{\braket{a e^{-i \pi \gamma}}_{n, \ell}}{\braket{a^{(0)}}_{n, \ell}} \xrightarrow{n \to \infty} b_\ell(s) \, ,\\ \label{identification_cft_sugra2}
\text{dDisc} \left[ \mathcal{H}(z, \zb)\right] \xrightarrow{\text{flat space}} \text{disc}\mathcal{A}_5 \, .
\end{gather}
Once again, these relations apply to the full double discontinuity and have been proven to be exact at order $\kappa=2$.
\subsection{The two loop example as the starting point}\label{sec:DiscAndCFT}
While we collect all the details for the $\kappa=3$ case in Appendix \ref{exp_res_ampl} and \ref{ampl_cft_comp}, here we would like to briefly discuss its most relevant features,  since this example clearly illustrates the two main points that will characterize the higher loop generalization, namely {\textit{the appearance of multi-trace operators}} in the OPE and the relation between leading logarithms and {\textit{iterated $s$-cuts}}.

Up to two loops, Eq.~\eqref{agamma} reads 
\begin{align}\label{comparisonCFT}
&\frac{\braket{a e^{-i \pi \gamma}}_{n, \ell}}{ \braket{a^{(0)}}_{n, \ell}}\to 1+ \frac{i \pi n^{3}}{2c(\ell+1)}+\frac{i \pi}{\ell+1} \int_0^1\frac{d\zb}{\zb^2}\left(\frac{1-\sqrt{1-\zb}}{1+\sqrt{1-\zb}}\right)^{\ell+1}\times \\ & \nonumber \qquad  \qquad \qquad \qquad\left\lbrace\frac{n^{11}}{c^2}g^{(2)}(\zb)+\frac{n^{19}}{c^3}\left(\log(1-\zb)g^{(3)}(\zb)+\underbrace{\cdots}_{\text{from} \log^2U}\right)+ \mathcal{O}\left(\frac{n^{8\kappa-5}}{c^{\kappa}}\right)\right\rbrace \, ,
\end{align}
where $g^{(2)}(\zb)$ is given in Eq.~(5.22) of \cite{Alday:2017vkk} and the dots represent the pieces  in the $\mathcal{H}^{(3)}$ correlator multiplying  $\log^2 (z \zb)$. The latter terms contribute to the dDisc,  but unfortunately we do not have an expression for them. There are both technical and conceptual problems to compute these terms, just to mention few of them, higher trace operators appear in the OPE and contribute to the dDisc and  the mixing problem needs to be solved at a higher $N$ order (e.g. $\gamma_{n, \ell, I}^{(2)}$ needs to be computed).\\
On the  supergravity side  we have that \eqref{bell} reduces to
\begin{align}\label{comparisonAmpl}
&b_\ell(s)=1+ \left(\frac{L \sqrt{s}}{2}\right)^3\frac{i \pi }{2c(\ell+1)}+\frac{i \pi}{\ell+1} \int_0^1\frac{d\zb}{\zb^2}\left(\frac{1-\sqrt{1-\zb}}{1+\sqrt{1-\zb}}\right)^{\ell+1}\times \\ & \nonumber  \qquad \qquad \qquad\left\lbrace \left(\frac{L \sqrt{s}}{2}\right)^{11}\frac{1}{c^2}g_2(\zb)+ \left(\frac{L \sqrt{s}}{2}\right)^{19}\frac{1}{c^3} \text{disc}_{x>1}F_3(x)\Big|_{x=\frac{1}{\zb}}+\mathcal{O}\left(\left(\frac{L \sqrt{s}}{2}\right)^{\kappa}c^{-\kappa}\right)\right\rbrace  \, ,
\end{align}
 where 
 \begin{align}
\text{disc}_{x>1}F_3(x) \equiv \frac{1}{2 \pi i}\left(F_3(x+i 0)-F_3(x-i 0) \right) \, .
\end{align} 
Comparing the two results, the first thing we notice is that given the form of dDisc$\mathcal{H}^{(3)}|_{\log^3 z \zb}$, only the planar diagrams contribute. The finite part of the non-planar one, in fact, has a completely different structure: the highest transcendental weight\footnote{Here and in the following we will consider as ``highest transcendental weight" those functions that have the highest transcendentality and with a non vanishing discontinuity, thus disregarding constant such as powers of $\pi$ or $\zeta_n$.} is three and consequently two for the discontinuity. As we are interested in matching the highest logarithmic part and this contribution is subleading, we can discard it in the comparison with the CFT. We believe that the non planar contributions will appear in the $\log^2 U$ and lower terms of the four-point correlator. Even when restricting to $I^{pl}_{db}$, Eqs.~\eqref{comparisonCFT} and \eqref{comparisonAmpl} do not agree completely. To investigate the motivation behind this mismatch, we decided to get back to the definition of a discontinuity and its interpretation in terms of \textit{unitarity cuts}.

It is know \cite{Cutkosky:1960sp} that the $s$-channel discontinuity of a generic amplitude $\mathcal{A}$, disc$_s\mathcal{A}$, is given by the sum over all possible cuts where $s$ is the momentum flowing.  A given cut diagram can be constructed from the integral representation by putting on-shell the cut propagators. In our case\footnote{At two loop and restricting $F_3$ to the  planar contributions only, disc$_s \mathcal{A}=-\text{disc}_{x>1}F_3(x)$.}, this translates diagrammatically in
\begin{equation} \label{cuts}
	2 \pi i \text{ disc}_s \mathcal{A} =\underbrace{
  \begin{tikzpicture}[baseline={([yshift=-1.1ex]current bounding box.center)},scale=0.28,node/.style={draw,shape=circle,fill=black,scale=0.4}]
   \draw [thick] (0,0) -- (6,0)--(6,3)--(0,3)--(0,0);
    \draw [thick] (3,0) -- (3,3);
    \draw [thick] (0,0) -- (-1,-1);
    \draw [thick] (0,3) -- (-1,4);
    \draw [thick] (6,0) -- (7,-1);
    \draw [thick] (6,3) -- (7,4);
    \draw [dashed,thick,red] (1.5,4)--(1.5,-1);
  \end{tikzpicture}+ \begin{tikzpicture}[baseline={([yshift=-1.1ex]current bounding box.center)},scale=0.28,node/.style={draw,shape=circle,fill=black,scale=0.4}]
   \draw [thick] (0,0) -- (6,0)--(6,3)--(0,3)--(0,0);
    \draw [thick] (3,0) -- (3,3);
    \draw [thick] (0,0) -- (-1,-1);
    \draw [thick] (0,3) -- (-1,4);
    \draw [thick] (6,0) -- (7,-1);
    \draw [thick] (6,3) -- (7,4);
    \draw [dashed,thick,red] (4.5,4)--(4.5,-1);
  \end{tikzpicture}}_{c_1}+\underbrace{\begin{tikzpicture}[baseline={([yshift=-1.1ex]current bounding box.center)},scale=0.28,node/.style={draw,shape=circle,fill=black,scale=0.4}]
   \draw [thick] (0,0) -- (6,0)--(6,3)--(0,3)--(0,0);
    \draw [thick] (3,0) -- (3,3);
    \draw [thick] (0,0) -- (-1,-1);
    \draw [thick] (0,3) -- (-1,4);
    \draw [thick] (6,0) -- (7,-1);
    \draw [thick] (6,3) -- (7,4);
    \draw [dashed,thick,red] (1.5,4)--(4.5,-1);
  \end{tikzpicture}+\begin{tikzpicture}[baseline={([yshift=-1.1ex]current bounding box.center)},scale=0.28,node/.style={draw,shape=circle,fill=black,scale=0.4}]
   \draw [thick] (0,0) -- (6,0)--(6,3)--(0,3)--(0,0);
    \draw [thick] (3,0) -- (3,3);
    \draw [thick] (0,0) -- (-1,-1);
    \draw [thick] (0,3) -- (-1,4);
    \draw [thick] (6,0) -- (7,-1);
    \draw [thick] (6,3) -- (7,4);
    \draw [dashed,thick,red] (4.5,4)--(1.5,-1);
  \end{tikzpicture}}_{c_2} \, .
\end{equation}
From this picture we can clearly separate two types of contributions
and we can try to reinterpret them from the CFT point of view,
similarly as done in  \cite{Bissi:2020wtv,Meltzer:2019nbs}.

We argue that the first type of cut ($c_1$) translates the double trace operator exchange, the
only one that we have information about, while the second one ($c_2$) corresponds
to the exchange of triple trace operators. The two contributions can be computed separately, thus allowing us
to compare each of them with the CFT. Naively one can think that $c_1$ should reproduce $g^{(3)}$ since we have concretely disentangled the triple trace contributions, however that is not the case. In fact, we have to remember that the conjecture connects the discontinuity of the amplitude, and relatedly $c_1$, with the entire dDisc and we are missing $\log^2 U$ term in the correlator. So we have to carefully rethink the results that we obtain when considering  the double discontinuity of the $\log^3 U$ term and interpret them from a different perspective. We know that when computing $\mathcal{H}^{(3)}|_{\log^3(z \zb)}$ we are taking the product of three tree-level anomalous dimensions $\gamma^{(1)}$, so the question one should ask is: what is the analogue of this quantity on the amplitude side? \\
In perturbative quantum field theory, one can 
compute the imaginary part of an amplitude at a generic loop $L$
knowing the results at lower loops, namely up to $(L-1)$. Similar
unitarity methods hold in CFT, see  \cite{Aharony:2016dwx,
  Meltzer:2019nbs, Meltzer:2020qbr} for example, where instead of
integrating over the phase space of the intermediate states, as one
usually does for amplitudes, one has to sum over all the states
exchanged in the OPE of the correlator. Reinterpreted from this
perspective, at two loops it was naive to expect that the results from
$c_1$, which knows intrinsically  about one loop information, would
have matched exactly the dDisc of the higher logarithmic part of the correlator,
constructed from $\left(\gamma^{(1)}\right)^3$. So to make a sensible
comparison, one should look for a different object built from the
amplitude and  depending only on tree-level ``two-particles" data.  In
\cite{Bissi:2020wtv}, we found that such an object exists and that the
CFT contribution we have computed is proportional to a saturated
$s$-channel cut, a diagram with all the possible $s$-cuts passing only
through two propagators. At this loop order, we call this quantity
\textit{double cut}, $c_{dc}$. Remarkably we find a perfect agreement, i.e.
\begin{align}\label{doublecutandg3}
\qquad \begin{tikzpicture}[baseline={([yshift=-1.1ex]current bounding box.center)},scale=0.25,node/.style={draw,shape=circle,fill=black,scale=0.4}]
   \draw [thick] (0,0) -- (6,0)--(6,3)--(0,3)--(0,0);
    \draw [thick] (3,0) -- (3,3);
    \draw [thick] (0,0) -- (-1,-1);
    \draw [thick] (0,3) -- (-1,4);
    \draw [thick] (6,0) -- (7,-1);
    \draw [thick] (6,3) -- (7,4);
    \draw [dashed,thick,red] (1.5,4)--(1.5,-1);
    \draw [dashed,thick,red] (4.5,4)--(4.5,-1);
  \end{tikzpicture} = 32 s^5 g^{(3)}(\zb)
\end{align}
with $g^{(3)}(\zb)$ as defined in Eq.~\eqref{g3}, where the matching is also with all the numerical factors. 
\subsection{Higher Loop generalisation}\label{sec_higher_loops}
In the spirit of understanding more and more how unitarity methods which are valid for S-matrix elements can be applied in the CFT context, we would like to extend the validity of this iterated $s$-cut construction to all loops and eventually compare these results with the ones in Sec.~\ref{sec_alllcft}. Even if our analysis will be restricted to flat space, similar arguments should remain valid for the full AdS$_5 \times$S$^5$ amplitude, thus possibly shedding some light on potential structures appearing there and how to extract them.

To actually compute this iterated cut, one can think of generalizing the differential equation
method we have been using for $\kappa=3$ to higher loops. However this becomes computationally hard already at
three loops and at higher order it is not possible to find a way to a priori fix the needed  boundary conditions. An alternative approach is thus necessary. Before illustrating it, let us comment that despite the possible appearance of higher order divergences as $\kappa$ increases, it has been known for a while and largely used in the literature that the iterated two-particle cut is actually finite to all orders \cite{Bern:2012uc}, as we have explicitly seen at two loops.

Let us start by going back to the general definition of a cut: a given cut diagram is a Feynman diagram in which the cut propagators are put on-shell. In the related integral, this corresponds to replace $\frac{1}{p^2-m^2+ i \epsilon}$ with $-2\pi i \delta^{(+)}(p^2-m^2)$, a one-dimensional delta function of the same argument, for each cut propagator. For example, in the one loop case, we have
\begin{equation} \label{sCutBox}
\begin{tikzpicture}[baseline={([yshift=-1.1ex]current bounding box.center)},scale=0.3,node/.style={draw,shape=circle,fill=black,scale=0.4}]
   \draw [thick] (0,0) -- (3,0)--(3,3)--(0,3)--(0,0);
    \draw [thick] (0,0) -- (-1,-1);
    \draw [thick] (0,3) -- (-1,4);
    \draw [thick] (3,0) -- (4,-1);
    \draw [thick] (3,3) -- (4,4);
    \draw [dashed,thick,red] (1.5,4)--(1.5,-1);
  \end{tikzpicture} \propto \int \text{d}^D k_1 \hspace{2mm} \frac{\delta^{(+)}(k_1^2)\delta^{(+)}((k_1-p_1-p_2)^2)}{(k_1-p_1)^2 (k_1+p_4)^2}  \, ,
\end{equation}
where $k_1$ represents the loop momentum. Generalized to  $(\kappa-1)$ loop, the iterated $s$-channel cut for the $\kappa$ rungs ladder diagram takes the form
\begin{align} \label{iteratedSch}
 \begin{tikzpicture}[baseline={([yshift=-1.1ex]current bounding box.center)},scale=0.3,node/.style={draw,shape=circle,fill=black,scale=0.4}]
   \draw [thick] (0,0) -- (6,0)--(6,3)--(0,3)--(0,0);
    \draw [thick] (3,0) -- (3,3);
    \draw [thick] (0,0) -- (-1,-1);
    \draw [thick] (0,3) -- (-1,4);
    \draw [thick] (6,0) -- (6.8,0);
    \draw [thick] (6,3) -- (6.8,3);
    \draw [thick,dashed] (7.2,0) -- (9.8,0);
    \draw [thick,dashed] (7.2,3) -- (9.8,3);
    \draw [thick] (9.3,0) -- (13,0);
    \draw [thick] (9.3,3) -- (13,3);
    \draw [thick] (10,0) -- (10,3);
    \draw [thick] (13,0) -- (13,3);
    \draw [thick] (13,0) -- (14,-1);
    \draw [thick] (13,3) -- (14,4);
     \node [below left=0.1 cm] at (-0.5,-0.5)  {$p_1$};
    \node [below right=0.1 cm] at (13.5,-0.5)   {$p_4$};
    \node [above right=0.1cm] at (13.5, 3.5) {$p_3$};
    \node [above left=0.1 cm] at (-0.5,3.5)  {$p_2$};
    \draw [dashed,thick,red] (1.5,4)--(1.5,-1);
    \draw [dashed,thick,red] (4.5,4)--(4.5,-1);
    \draw [dashed,thick,red] (11.5,4)--(11.5,-1);
  \end{tikzpicture} \propto \int \prod_{i=1}^{\kappa-1} \text{d}^D k_i \frac{\delta^{(+)}(q_1^2)\delta^{(+)}(q_2^2) \cdots \delta^{(+)}(q_{2\kappa-2}^2)}{q^2_{2 \kappa-1} \cdots q^2_{3\kappa-2}} \, ,
\end{align}
where $q_1^{\mu}$ is a combination of external and loop momenta $k_i^{\mu}$. Thus another way to compute the iterated $s$-channel cut is to solve directly this integral. Before continuing with the details of the computation, let us set some notation: we are working in Lorentzian signature (diag$(1, -1, \ldots)$), in our conventions all the $D$-dimensional external momenta are outgoing and in the centre of mass frame they take the following form
\begin{equation*}
p_1^{\mu}=\frac{1}{2 \sqrt{s}}\{1,1, \vec{0}_{D-2}\}\, , \quad p_2^{\mu}=\frac{1}{2 \sqrt{s}}\{1,-1, \vec{0}_{D-2}\}\, , \quad p_4^{\mu}=\frac{1}{2 \sqrt{s}}\{-1,-\cos\chi,-\sin\chi, \vec{0}_{D-3}\} \, .
\end{equation*}
Energy-momentum conservation fixes $p_1^{\mu}+p_2^{\mu}+p_3^{\mu}+p_4^{\mu}=0$. The loop momenta can be parametrized as
\begin{equation}
k_1^{\mu}= \frac{1}{2 \sqrt{s}} \{E_{k_1},|k_1|\cos\theta_1,|k_1| \sin\theta_1 \cos\theta_2, |k_1| \sin\theta_1\sin\theta_2,\vec{0}_{D-4} \}\,,
\end{equation}
and the corresponding measure
\begin{equation}
\int \text{d}^D k_1 \propto \int \text{d}E_{k_1}   \int \text{d}|k_1| |k_1|^\frac{D-1}{2}  \int_0^{2 \pi} \text{d}\theta_2 \hspace{2mm} \sin^{D-4}\theta_2  \int_0^\pi \text{d}\theta_1 \hspace{2mm} \sin^{D-3}\theta_1\,.
\end{equation}
From now on we will fix $D$ to be $10$ and we will start by computing the $s$-cut of the one loop box integral in Eq.~\eqref{sCutBox}. The two delta functions can be used to perform the integration in $E_{k_1}$ and $|k_1|$, such that the loop momentum reduces to
\begin{equation}
k_1^{\mu \, OS}= \frac{1}{2 \sqrt{s}} \{1,\cos\theta_1, \sin\theta_1 \cos\theta_2, 
\sin\theta_1\sin\theta_2,\vec{0}_{D-4} \}\,.
\end{equation}
After this simplification, the integral \eqref{sCutBox} becomes\footnote{Similar results were already know in the literature, see for example Appendix A of \cite{vanNeerven:1985xr}. AG thanks S. Abreu for pointing out this reference.}
\begin{equation}\label{10Dcos}
\int \underbrace{\text{d}\theta_2 \, \text{d}\theta_1  \sin^6\theta_2 \sin^7\theta_1}_{\text{measure}}\underbrace{\frac{1}{s^2(1-\cos\theta_1)}}_{I_{left}}\times \underbrace{\frac{1}{(1-\cos\chi \sin\theta_1- \cos\theta_2 \sin\chi \sin\theta_1)}}_{I_{right}}   \, , 
\end{equation}
where we have suppressed some numerical factors for simplicity and beside the measure, we can distinguish two pieces: a left and a right integrand. The $\theta_2$ integration can be done straightforwardly, then we can perform a simple change of variable, $\cos\theta_1= 2 v-1$, to express the remaining integral as one in $v$ ranging from 0 to 1. To make contact with the usual expressions in terms of Mandelstam invariants, we can identify $\cos \chi=\frac{s+2t}{s}$. The integrand then takes the form
\begin{align}\label{kernel}
&\frac{\pi }{2 s t^3 (v-1) (s+t)^3}(-s (s+t-s v )^4 \left|
   \frac{t}{s}-v+1\right| +(s+t)^5-5 s v (s+t)^4\\& \nonumber +10 s^2 v^2 (s+t)^3-10 v^3 (s+t)^2 \left(s^3+2 t^3\right)+5 v^4 (s+t) \left(s^4+8 s t^3+6 t^4\right)\\& \nonumber 
 +v^5 \left(-20 s^2 t^3-s^5-30 s t^4-12
   t^5\right)) \equiv \frac{\mathcal{K}(v, s, t)}{2s(1-v)} \, .
\end{align}
The absolute value comes from the square roots appearing after the $\theta_2$ integration and as a consequence of their presence the integration region is split in  two subsets: $v$ less or greater than $\frac{s+t}{s}$ (where we are considering $s<0$). 
The last integration then gives
\begin{equation}
\mathcal{R}_1(s,t)=\begin{tikzpicture}[baseline={([yshift=-1.1ex]current bounding box.center)},scale=0.3,node/.style={draw,shape=circle,fill=black,scale=0.4}]
   \draw [thick] (0,0) -- (3,0)--(3,3)--(0,3)--(0,0);
    \draw [thick] (0,0) -- (-1,-1);
    \draw [thick] (0,3) -- (-1,4);
    \draw [thick] (3,0) -- (4,-1);
    \draw [thick] (3,3) -- (4,4);
    \draw [dashed,thick,red] (1.5,4)--(1.5,-1);
  \end{tikzpicture} \propto \int_0^{\frac{t+s}{s}} \text{d}v \hspace{2mm} \frac{\mathcal{K}^+(v,s,t)}{2 s (1-v)}+\int_{\frac{t+s}{s}}^1 \text{d}v \hspace{2mm} \frac{\mathcal{K}^-(v,s,t)}{2 s (1-v)}  \,, 
\end{equation}
where we have defined 
\begin{equation}\label{eq:Kpm}
\mathcal{K}^\pm = \left. \mathcal{K}\right|_{|x|=\pm x}\,.
\end{equation}
This matches the result of \cite{Alday:2017vkk} up to an overall factor which comes from normalization.

The uptake of this construction is that it can be easily generalized to higher loops. Similarly to the box integral which was constructed integrating over the product of the two pieces in Eq.~\eqref{10Dcos} separated by the cut, we can now glue the same right integrand to the one-loop result to construct the two-loop case. Then the idea is that to compute the $L$-loop integral, we can keep on multiplying the right integrand in  Eq.~\eqref{10Dcos} for the result obtained at order $(L-1)$.\\
Let us start from the two-loop case and let us see how we can retrieve the result  in Sec.~\ref{sec:DiscAndCFT} for the \textit{double cut}. Notice that the left integrand in Eq.~\eqref{10Dcos} has no $\theta_2$ dependence, so we can directly take the integrated result \eqref{kernel} and multiply it by the left integrand, i.e. $2 s (1-v)$, to obtain the needed kernel $\mathcal{K}(v,s,t)$. Consequently, at two loops we get
\begin{equation}\label{twoloopiterated}
 \begin{tikzpicture}[baseline={([yshift=-1.1ex]current bounding box.center)},scale=0.25,node/.style={draw,shape=circle,fill=black,scale=0.4}]
   \draw [thick] (0,0) -- (6,0)--(6,3)--(0,3)--(0,0);
    \draw [thick] (3,0) -- (3,3);
    \draw [thick] (0,0) -- (-1,-1);
    \draw [thick] (0,3) -- (-1,4);
    \draw [thick] (6,0) -- (7,-1);
    \draw [thick] (6,3) -- (7,4);
    \draw [dashed,thick,red] (1.5,4)--(1.5,-1);
    \draw [dashed,thick,red] (4.5,4)--(4.5,-1);
  \end{tikzpicture} \propto \int_0^{\frac{t+s}{s}} \text{d}v \hspace{1mm} \mathcal{R}_1(s,-s \hspace{0.2mm} (1-v) ) \mathcal{K}^+(v,s,t)+\int_{\frac{t+s}{s}}^1 \text{d}v \hspace{1mm}  \mathcal{R}_1(s,-s \hspace{0.2mm}(1-v) ) \mathcal{K}^-(v,s,t)\, ,
\end{equation}
where the new definition of $t$ comes from the fact that the leg
associated to $p_4^{\mu}$ is now $k_1^{\mu\,OS}$. This result matches both with the CFT $g^{(3)}(\zb)$ in Eq.~\eqref{g3} and with the cut constructions made using differential equations. With respect to the latter, the computation presented here has the advantage of being faster and more
efficient as it does not  require the knowledge of a canonical basis or
integral by parts identities.\\
As an aside comment, let us mention that at two loop we can obtain $c_1$ through a similar construction, the only difference is that now we have to use  one-loop
integral as the right integrand. To be more precise,
\begin{equation}
 \begin{tikzpicture}[baseline={([yshift=-1.1ex]current bounding box.center)},scale=0.25,node/.style={draw,shape=circle,fill=black,scale=0.4}]
   \draw [thick] (0,0) -- (6,0)--(6,3)--(0,3)--(0,0);
    \draw [thick] (3,0) -- (3,3);
    \draw [thick] (0,0) -- (-1,-1);
    \draw [thick] (0,3) -- (-1,4);
    \draw [thick] (6,0) -- (7,-1);
    \draw [thick] (6,3) -- (7,4);
    \draw [dashed,thick,red] (1.5,4)--(1.5,-1);
  \end{tikzpicture} \propto \int_0^{\frac{t+s}{s}} \text{d}v
  \hspace{1mm} \begin{tikzpicture}[baseline={([yshift=-1.1ex]current bounding box.center)},scale=0.3,node/.style={draw,shape=circle,fill=black,scale=0.4}]
   \draw [thick] (0,0) -- (3,0)--(3,3)--(0,3)--(0,0);
    \draw [thick] (0,0) -- (-1,-1);
    \draw [thick] (0,3) -- (-1,4);
    \draw [thick] (3,0) -- (4,-1);
    \draw [thick] (3,3) -- (4,4);
  \end{tikzpicture} 
    \mathcal{K}^+(v,s,t)+\int_{\frac{t+s}{s}}^1 \text{d}v
  \hspace{1mm} \begin{tikzpicture}[baseline={([yshift=-1.1ex]current bounding box.center)},scale=0.3,node/.style={draw,shape=circle,fill=black,scale=0.4}]
   \draw [thick] (0,0) -- (3,0)--(3,3)--(0,3)--(0,0);
    \draw [thick] (0,0) -- (-1,-1);
    \draw [thick] (0,3) -- (-1,4);
    \draw [thick] (3,0) -- (4,-1);
    \draw [thick] (3,3) -- (4,4);
  \end{tikzpicture}  \mathcal{K}^-(v,s,t) \,, 
\end{equation}
where again we have redefined $t$ to be $-s \hspace{0.2mm}
(1-v)$. Also in this case the
integration is finite, as the $c_1$ divergence comes only from the box
contribution. For this reason we have to compute the kernel $\mathcal{K}(v,s,t)$ up to order
$\epsilon$ to obtain the finite part of the result.
Fortunately this can be done, it is sufficient to perform the $\theta_2$ integration in
$d-$dimension, which gives us a more involved result containing ${}_2
F_1$ functions, and then expand the result up to the required
order.

As anticipated before, the procedure yielding to Eq.~\eqref{twoloopiterated} can be  easily extended to all loops as
\begin{equation}\label{alllopsampl}
\mathcal{R}_L(s,t)= \int_0^{\frac{t+s}{s}} \text{d}v \hspace{1mm} \mathcal{R}_{L-1}(s,-s \hspace{0.2mm}(1-v)) \mathcal{K}^+(v,s,t)\hspace{1mm} +\int_{\frac{t+s}{s}}^1 \text{d}v \hspace{1mm} \mathcal{R}_{L-1}(s,-s \hspace{0.2mm} (1-v)) \mathcal{K}^-(v,s,t) \,.
\end{equation}
With these results at hand we can now finally compare them with the ones obtained in Sec.~\ref{sec:FSLeadingLogs}: we find matches up to $20-$loops for the highest
weight coefficients (see Appendix \ref{appendix:polStructure}) and up to $6-$loops for the full answer.  This  is a highly non trivial check  of the method presented above and of our  conjecture relating the dDisc of leading logarithmic terms in the correlator, up to a log factorization\footnote{To have an exact match between iterated cuts and CFT correlator in the flat space limit without any log factorization, one can consider instead of the double discontinuity some sort of $\underbrace{\text{d}\,\ldots\,\text{d}}_{\kappa -1}$Disc, such that it returns directly  $g^{(\kappa)}(\zb)$.},  and iterated $s$-channel cuts of the amplitude. Pictorially this translates in
\begin{align}
 \begin{tikzpicture}[baseline={([yshift=-1.1ex]current bounding box.center)},scale=0.25,node/.style={draw,shape=circle,fill=black,scale=0.4}]
   \draw [thick] (0,0) -- (6,0)--(6,3)--(0,3)--(0,0);
    \draw [thick] (3,0) -- (3,3);
    \draw [thick] (0,0) -- (-1,-1);
    \draw [thick] (0,3) -- (-1,4);
    \draw [thick] (6,0) -- (6.8,0);
    \draw [thick] (6,3) -- (6.8,3);
    \draw [thick,dashed] (7.2,0) -- (9.8,0);
    \draw [thick,dashed] (7.2,3) -- (9.8,3);
    \draw [thick] (9.3,0) -- (13,0);
    \draw [thick] (9.3,3) -- (13,3);
    \draw [thick] (10,0) -- (10,3);
    \draw [thick] (13,0) -- (13,3);
    \draw [thick] (13,0) -- (14,-1);
    \draw [thick] (13,3) -- (14,4);
    \draw [dashed,thick,red] (1.5,4)--(1.5,-1);
    \draw [dashed,thick,red] (4.5,4)--(4.5,-1);
    \draw [dashed,thick,red] (11.5,4)--(11.5,-1);
  \end{tikzpicture} \propto g^{(\kappa)}(\zb)
\end{align}
where $g^{(\kappa)}(\zb)$ is defined in Eq.~\eqref{FlatSpaceh}.\\
Interestingly the same approach can be in principle used in any dimension, except $4$ where the interested integral is actually divergent.

{Before concluding, let us mention that this identification can also be interpreted as an a posteriori justification for our choice of focusing only on planar ladder diagrams when dealing with CFT leading logarithm terms. These diagrams are indeed the only ones presenting exactly $(\kappa -1)$  iterated two-particle cuts and as a consequence containing the same information as in the CFT leading log pieces. In going to higher order one should be more careful  on the type of integrands appearing in the amplitude and the associated Feynman diagrams but we believe that all other diagrams apart from planar ladders should map to different $\log^k U \log^j V$, with $k$ strictly less than $\kappa$.}
For example at three loops we believe that 
\begin{equation}
\begin{tikzpicture}[baseline={([yshift=-1.1ex]current bounding box.center)},scale=0.18,node/.style={draw,shape=circle,fill=black,scale=0.4}]
   \draw [thick] (0,0) -- (6,0)--(6,6)--(0,6)--(0,0);
    \draw [thick] (3,0) -- (3,6);
    \draw [thick] (3,3) -- (6,3);
    \draw [thick] (0,0) -- (-1,-1);
    \draw [thick] (0,6) -- (-1,7);
    \draw [thick] (6,0) -- (7,-1);
    \draw [thick] (6,6) -- (7,7);
  \end{tikzpicture} \approx  \log^3(U) \log^3(V)
\end{equation}
{Having a complete understanding of higher loops contributions enabling us to check our intuitions, would also require the knowledge of the unmixed OPE data starting already from the ones at order $c^{-1}$.  This represents a very challenging problem, so for the moment our discussion for $\kappa \geq 4$ should be taken as a collection of qualitative  observations  based on explicit examples.  We plan nonetheless to give a more concrete answer to these problems in the future.}
\section{Mellin space}\label{sec:Mellin}
There are certain circumstances, such as in the direct comparison with
a gravity amplitude, in which certain properties and
characteristics become more visible and accessible when considering
the Mellin representation of CFT correlators. In this section we will
reinterpret our previous results from this prospective in an
attempt to understand more deeply the physics in the flat space limit.
\subsection{Generalities and flat space limit}\label{sec:Mell2l}
In Mellin space $\mathcal{H}(U,V)$ is defined through the integral \cite{Mack:2009mi, Rastelli:2016nze, Rastelli:2017udc}
\begin{align} \label{MellTranform}
\mathcal{H}(U,V)=\int_{-i \infty}^{i \infty} \frac{\text{d}s \text{d}t}{(4 \pi i)^2}U^{s/2}V^{(t-4)/2}\mathcal{ \widetilde{M}}(s,t)\Gamma\left(\frac{4-s}{2} \right)^2\Gamma\left(\frac{4-t}{2} \right)^2\Gamma\left(\frac{4-u}{2} \right)^2 \, ,
\end{align}
where $s, \, t$ and $u$ are dependent variables, very reminiscent of Mandelstam ones, satisfying the constraint $s+t+u=4$ and $\mathcal{ \widetilde{M}}(s, t)$ is  the (\textit{reduced}) \textit{Mellin Amplitude}. In this language, the crossing equations \eqref{eq:crossing} translates into
\begin{align}
\mathcal{ \widetilde{M}}(s, t)=\mathcal{ \widetilde{M}}(t, s) \, .
\end{align}
The Mellin amplitude inherits from the correlator in the supergravity limit  a loop expansion around large $c$. A term in $\mathcal{H}^{(\kappa)}(U, V)$ that behaves as $\sim \log^k U \log^j V$ would be generated in Mellin space from an infinite sum over simultaneous poles in $s$ and $t$ of the form $\sum_{m, n}\nu_{mn}(s-2m)^{-k+1}(t-2n)^{-j+1}$. In particular it has been shown in \cite{Alday:2018kkw} that at one loop
\begin{align} \label{MellOneLoop}
\mathcal{\widetilde{M}}^{(2)}(s, t)=\sum_{m,n=2}^{\infty} \left( \frac{c_{mn}^{(2)}}{(s-2m)(t-2n)}+\frac{c_{mn}^{(2)}}{(t-2m)(u-2n)}+\frac{c_{mn}^{(2)}}{(u-2m)(s-2n)} \right)
\end{align}
it is sufficient to successfully reproduce the entire $\mathcal{H}^{(2)}(U, V)$, without the addition of any other simple pole or regular term.

The Mellin space formalism turns out to be mostly suited when studying scattering amplitudes. In fact, it has been argued in \cite{Penedones:2010ue, Fitzpatrick:2011ia,Paulos:2011ie} that scattering amplitudes in AdS are Mellin transform of CFT correlators and that in the large $s, \,t$ limit $\mathcal{\widetilde{M}}(s, t)$ recovers exactly the amplitude computed in flat  space-time. The prescription to take this flat space limit, or equivalently large AdS radius, is \cite{Alday:2018pdi,Chester:2018dga}
\begin{align}\label{FSMell}
\lim_{L \to \infty} L^{14} \pi^3 \frac{\Theta_4^{flat}(s, t, \sigma, \tau)}{16} \mathcal{\widetilde{M}}\left(L^2 s, L^2 t \right)=\int_{0}^{\infty} d\beta \beta e^{-\beta} \mathcal{A}_{10}^{\perp} \left(2\beta s, 2\beta t , \sigma, \tau\right) \, ,
\end{align}
where $\sigma$ and $\tau$ are the $SU(4)_R$ cross-ratios in Eq.~\eqref{crossRatios}
 and $\Theta_4^{flat}(s, t, \sigma, \tau)\equiv (t u +t s\sigma+s u \tau)^2$, see \cite{Alday:2018pdi} and references therein for more details on the derivation of this formula. With $\mathcal{A}_{10}^{\perp}$ we denote the ten dimensional flat space graviton amplitude in the transverse kinematics: the momenta $k_i$ lie on $\mathbb{R}^5 \simeq \text{AdS}_5|_{L \to \infty}$ and are orthogonal to the S$^5$ polarization vectors ($k_1 \cdot y_i=0 $). In this configuration $\mathcal{A}_{10}^{\perp}$ coincides with $\mathcal{A}_{10}^{sugra}$ in Eq.~\eqref{10dAmpl} with $\hat{K}=64 \, \Theta_4^{flat}(s,t, \sigma, \tau) c(2) \beta^4$ and $c(2)=\sfrac{1}{32 \pi^2}$.
\subsection{Two loops and beyond} 
 Given the behaviour of $\mathcal{H}^{(\kappa)} |_{\log^{\kappa}(z \zb)}$ in Eq.~\eqref{leadLogUV}, for the corresponding Mellin amplitude we consider
 \begin{align}\label{MellGenericK}
 \widetilde{\mathcal{M}}^{(\kappa)}_{\log} (s, t)= \sum_{m,n+2}^{\infty} \frac{c_{mn}^{(\kappa)}}{(s-2m)^{\kappa-1}(t-2n)} \, ,
 \end{align}
where the subscript $\log$ reminds us that we are restricting to the leading logarithmic terms of the correlation function and where we have reported  only one channel, all possible cyclic permutations have to be taken into account to get the full amplitude.\\
The explicit analysis at two loops, whose details can be found in Appendix \ref{Appendix:Mell}, has allowed us to formulate the following observations.
\begin{itemize}
\item The residue  integral associated to the terms in Eq.~\eqref{MellGenericK} produces not only $\log^\kappa U \log^2 V$ contributions but also lower $\log$ powers. For the case $\kappa=3$ for instance,
\begin{align}
\label{Ma}
\mathcal{H}^{(3)}(U, V)=\sum_{m,n=2}^{\infty}&\frac{(n-1)_{m+1}^2 c_{mn}}{24 \Gamma(m-1)^2}U^m V^{n-2} \log^3 U \log^2 V+\\ \nonumber & \frac{
   (n-1)_{m+1}^2 c_{mn}}{4 \Gamma (m-1)^2}U^m V^{n-2} \left(H_{m+n-1}-H_{m-2}\right) \log^2 U \log^2 V+ \cdots
\end{align}
where $H_l\equiv \sum_{j=1}^l \frac{1}{j}$ represents the Harmonic number. The dots represent contributions from lower powers of $\log U, \, \log V$ terms and from the other pieces of the correlator that we do not know explicitly.\\
From this expression, it seems that the knowledge of $\sim s^{1-\kappa} t^{-1}$ terms allow to infer something about lower order poles.  A deeper understanding of this  interplay  may give further insight on the full structure of the four-point function. This observation can be related to a similar effect that appears in the flat-space amplitude, where we can indeed reconstruct, up to terms that vanish after iterated discontinuities, the full double trace contribution.
\item  Now specifying to the two-loop case, in order to get exactly  $\mathcal{H}^{(3)}|_{\log^3(z \zb)}(U,V)$ , we have to consider
\begin{align}
\sum_{m,n=2}^{\infty} \left( \frac{c_{mn}}{(s-2m)^2 (t-2n)}+\frac{c_{mn}}{(s-2m)^2 (u-2n)} \right)\, .
\end{align}
Notice that  analogously to what happens in  the one-loop case,  there is no need to add any $\frac{1}{(s-2m)^2}$ pole. The other permutations appearing in $ \widetilde{\mathcal{M}}^{(3)}_{\log} (s, t) $, with $m$ and $n$ properly recombined, give crossing symmetric versions of the correlator. This last fact is in accordance with the discussion  in Appendix \ref{ampl_cft_comp}, where the dDisc of \eqref{correlator} is compared with only one channel of the amplitude.
\item The flat space limit of  $\widetilde{\mathcal{M}}^{(3)}_{\log} (s, t)$, concretely implemented taking $s, \, t \to \infty$ with $m$ and $n$ of the same order and carefully regularizing the resulting sum, surprisingly produces a $c_{mn}^{(3)}$ with the same polynomial structure as the one appearing in  $I^{pl}_{db}(s, t)$ in Eq.~\eqref{resDBPl}. However the functions multiplying these polynomials do not have the right transcendental weight to exactly reproduce the space-time amplitude. 
\item If one further studies the relations between the Mellin result and  $c_1$ and $c_2$ in Sec.~\ref{sec:DiscAndCFT}, it seems to exist a connection between logarithmic singularities, and consequently inverse powers of $(s-2m)$ and $(t-2n)$, and discontinuities, corroborating the intuition we have been following throughout the paper.
\end{itemize}
These observations, together with the identification of leading logs with iterated $s$-cuts, suggested us that the $c_{mn}^{(\kappa)}$'s should have an interpretation in terms of unitarity cuts. Driven buy this intuition, we get back to the easiest example, namely the one-loop Mellin amplitude in its flat space limit. To get it, we need to  take $m, \,n$ large and of order $\mathcal{O}(s,\, t)$, in this regime the sums become integrals and if we ignore for a second that these sums are divergent, this formula is very reminiscent of the Mandelstam representation for amplitudes\footnote{Amplitudes of spinless massive particles under certain specific assumptions can be expressed through the following representation:
\begin{align*}
\mathcal{A}(s,t)=\text{poles}+\frac{1}{\pi^2} \iint \frac{\rho_{st}(s^{\prime},t^{\prime})}{(s^{\prime}-s)(t^{\prime}-t)}ds^{\prime} dt^{\prime}+\frac{1}{\pi^2} \iint \frac{\rho_{tu}(t^{\prime},u^{\prime})}{(t^{\prime}-t)(u^{\prime}-u)}dt^{\prime} du^{\prime}+\frac{1}{\pi^2} \iint \frac{\rho_{su}(s^{\prime},u^{\prime})}{(s^{\prime}-s)(u^{\prime}-u)}ds^{\prime} du^{\prime}
\end{align*}
where $\rho_{xy}=\text{disc}_x\text{disc}_y \mathcal{A}$ is called the double spectral density.} \cite{Eden:1966dnq,Correia:2020xtr}. Hence, it is tempting to compare the $c_{mn}^{(2)}$ with $\text{disc}_s \text{disc}_t I_{box}(s, t)$ and indeed one obtains
\begin{align} 
\begin{tikzpicture}[baseline={([yshift=-1.1ex]current bounding box.center)},scale=0.4,node/.style={draw,shape=circle,fill=black,scale=0.4}]
   \draw [thick] (0,0) -- (3,0)--(3,3)--(0,3)--(0,0);
    \draw [thick] (0,0) -- (-1,-1);
    \draw [thick] (0,3) -- (-1,4);
    \draw [thick] (3,0) -- (4,-1);
    \draw [thick] (3,3) -- (4,4);
    \draw [dashed,thick,red] (1.5,4)--(1.5,-1);
    \draw [dashed,thick,red] (4,1.5)--(-1,1.5);
  \end{tikzpicture} \propto c_{mn}^{(2)} \Big|_{m=\frac{s}{2},n=\frac{t}{2}}=\frac{3 \, s^2 t^2}{(s+t)^3} \, ,
\end{align}
where the factor of proportionality is fixed by the $\beta$ integral in Eq.~\eqref{FSMell}. At one loop the $c_{mn}^{(2)}$ are obtained studying $\log^2 U \log^2 V$ term in the correlation function, so what we get seems in accordance with the identification of the double spectral density $\rho_{st}$ and CFT qDisc \cite{Correia:2020xtr}, which selects exactly this term. The qDisc of a correlator is defined as the quadruple discontinuity obtained by taking one double discontinuity in $\zb=1$, followed by a dDisc around $z=0$, in this way one focuses on terms which have at least a $\log^2 (1-\zb)$ and a $\log^2 z$. Performing a similar analysis at two loops, we find\footnote{At a first glance, this should be surprising, since it is involved a ``triple-trace" type of cut, however a more careful analysis shows that $c_2$ in Eq.~\eqref{cuts} does not really  contribute.} 
\begin{align}\label{2loopsMdc}
\begin{tikzpicture}[baseline={([yshift=-1.1ex]current bounding box.center)},scale=0.3,node/.style={draw,shape=circle,fill=black,scale=0.4}]
   \draw [thick] (0,0) -- (6,0)--(6,3)--(0,3)--(0,0);
    \draw [thick] (3,0) -- (3,3);
    \draw [thick] (0,0) -- (-1,-1);
    \draw [thick] (0,3) -- (-1,4);
    \draw [thick] (6,0) -- (7,-1);
    \draw [thick] (6,3) -- (7,4);
    \draw [dashed,thick,red] (1.5,4)--(1.5,-1);
    \draw [dashed,thick,red] (4.5,4)--(4.5,-1);
    \draw [dashed,thick,red] (7,1.5)--(-1,1.5);
  \end{tikzpicture}\propto c_{mn} \Big|_{m=\frac{s}{2},n=\frac{t}{2}}=-\frac{3 t^5}{256} {}_2F_1\left(	D-4, D-4, \frac{3}{2}(D-4);-\frac{t}{s} \right)\Big|_{D=10} \, .
 \end{align}
We conjecture that the same should happen at higher loops, i.e. that in flat space the $c_{mn}^{(\kappa)}$'s  in Eq.~\eqref{MellGenericK} reproduce the \textit{maximal cut} of the corresponding $\kappa$ rungs ladder diagrams, where the \textit{maximal cut} corresponds to the  diagram with  all the propagators on-shell. Following similar ideas as in Sec.~\ref{sec_higher_loops} we can construct the maximal cut contribution for the ladder diagram as an iterated integral. In order to do so we apply the loop-by-loop construction of the Baikov representation \cite{Frellesvig:2017aai} obtaining\footnote{The analysis is dimensional independent but here we present the explicit results in ten dimensions.}
\begin{align}
c_{mn}^{(\kappa)} \Big|_{m=\frac{s}{2},n=\frac{t}{2}} \sim  \begin{tikzpicture}[baseline={([yshift=-1.1ex]current bounding box.center)},scale=0.3,node/.style={draw,shape=circle,fill=black,scale=0.4}]
   \draw [thick] (0,0) -- (6,0)--(6,3)--(0,3)--(0,0);
    \draw [thick] (3,0) -- (3,3);
    \draw [thick] (0,0) -- (-1,-1);
    \draw [thick] (0,3) -- (-1,4);
    \draw [thick] (6,0) -- (6.8,0);
    \draw [thick] (6,3) -- (6.8,3);
    \draw [thick,dashed] (7.2,0) -- (9.8,0);
    \draw [thick,dashed] (7.2,3) -- (9.8,3);
    \draw [thick] (9.3,0) -- (13,0);
    \draw [thick] (9.3,3) -- (13,3);
    \draw [thick] (10,0) -- (10,3);
    \draw [thick] (13,0) -- (13,3);
    \draw [thick] (13,0) -- (14,-1);
    \draw [thick] (13,3) -- (14,4);
    \draw [dashed,thick,red] (1.5,4)--(1.5,-1);
    \draw [dashed,thick,red] (4.5,4)--(4.5,-1);
    \draw [dashed,thick,red] (11.5,4)--(11.5,-1);
    \draw [dashed,thick,red] (-1,1.5)--(14,1.5);
  \end{tikzpicture} \propto \int_0^{t/s} \text{d}Z \left. c_{mn}^{(\kappa-1)} (s,t) \right|_{t=s Z}  \frac{F_b^{5}(s,t,Z)}{ G^3(s,t)}\,,
\end{align}
where we have defined the Gram determinant and the cut Baikov kernel as:
\begin{equation}
G=\frac{1}{4} s t (s+t)   \qquad F_b=\frac{1}{4} s (t - s Z) \,.
\end{equation}
We have also tested that the contribution of this maximal cut, after having performed the integration associated with the discontinuity in $t$, matches up to polynomial terms the one obtained with \eqref{alllopsampl}.

\section*{Acknowledgements}
We thank F. Alday and E. Perlmutter for several discussions and insights. AG also thanks
B. Page, S. Abreu, B. Basso, E. Trevisani, V. Goncalves, R. Pereira,
G. Korchemsky for useful discussion. 
The work of AB and GF is supported by Knut and Alice Wallenberg Foundation under grant KAW 2016.0129 and by VR grant 2018-04438.
The work of AG is supported by the Knut and Alice Wallenberg Foundation under grant 2015.0083 and by the French National Agency for Research grant ANR-17-CE31-0001-02 and ANR-17-CE31-0001-01.

\newpage

\appendix
\section{Harmonic Polylogarithms} \label{sec:HarmonicPolylog}
Harmonic Polylogarithms \cite{Remiddi:1999ew,Maitre:2005uu,Maitre:2007kp} are a generalization of the classical polylogarithms and similarly can be defined through some iterated integrations. We will denote with $H_{\cdots}(x)$ the harmonic polylogarithm of argument $x$ and transcendental weight $w$, where the dots represent a collection of $w$ indices. Starting from weight one,
\begin{align}
\HPL{0}{x}=\log(x) \quad \HPL{1}{x}=-\log(1-x) \quad \HPL{-1}{x}=\log(1+x) \, ,
\end{align}
we can generate all the HPL's of higher transcendentality as
\begin{equation}
H_{\,a \cdots}(x)= \int_0^x \text{d}x^\prime H_{\cdots}(x^\prime) \phi(a,x^\prime) 
\end{equation}
with
\begin{equation}
\phi(0,x)=\frac{1}{x}  \quad \phi(1,x)=\frac{1}{1-x} \quad \phi(-1,x)=\frac{1}{1+x} \, .
\end{equation}
From this definition it is clear that the usual polylogarithms  takes the form Li$_{n}(x)=H_{\vec{0}_{n-1} 1}(x)$.\\
These new functions present lots of interesting properties and advantages, one among others is that it is easy to study their analytic behaviour. In particular, HPL's with only $0$'s and $1$'s in their index vector, can develop  only two types of  logarithmic singularity, one at $x=0$ and the other one at $x=1$, that can be directly detected looking at the index vector itself. Divergences of type $\log^n x$ arises when there are $n$ 0's to the right end of the weight vector, while if $n$ 1's appear to the very left, the HPL will behave as $\log^n(1-x)$ for $x \to 1$. The ease of extracting singular behaviour will be really useful when comparing  CFT computations and amplitude results.
\section{The two loop example} \label{App:2loop}
In this Appendix we will report in more detail the results in  \cite{Bissi:2020wtv} for the two-loop example. We will first give the full answer for the highest logarithmic term of the correlation function of four identical $\mathcal{O}_2$ operators at order $c^{-3}$, then we will continue with a discussion of the two-loop amplitude, giving full results for the planar and non planar double box integrals. We continue with some comments on the  comparison between these two results. In the last section, we perform a parallel two-loop analysis in Mellin space. We refer to the ancillary Mathematica file for the longer expressions.
\subsection{Results for $\mathcal{H}^{(3)}|_{\log^3 (z \zb)}$}
\label{sec:2loopEx}
Let us start from the the expression for $h^{(3)}(z)$, according to Eq.~\eqref{eq:hsmall} and \eqref{pandqPol}, it takes the form
\begin{align}
h^{(3)}(z)&=p_3\left( \frac{z}{z-1} \right)\left(\HPL{001}{z}+\HPL{101}{z}\right)+p_3(z)\left(\HPL{001}{z}+\HPL{011}{z} \right)\\ & \nonumber +  q_3(z) \HPL{01}{z}+q_3\left( \frac{z}{z-1} \right)\left(\HPL{01}{z}+\HPL{11}{z}\right)+j_1(z)\HPL{1}{z}+j_0(z)\\ \label{p3}
p_3(z)&=-\frac{\left(10z^2+10 z+1\right)}{2880 z^5} \\ \label{q3}
q_3(z)&= \frac{1260 z^4-3870 z^3+4170 z^2-1785 z+227}{172800 z^5}\\ \label{j1}
j_1(z)&=\frac{(1258 z^3-4161 z^2   + 5806z-2903)}{172800 z^4}\\
j_0(z)&=-\frac{217855 z^3-685424  z^2+749142 z-499428}{12441600 z^3}
\end{align}

Now applying Eq.~\eqref{Hwithh}, we can extract\footnote{{To adapt to our convention we have restored the factor $(z \zb)^2$ and we have multiplied everything by $4^3 \cdot 2 \cdot 3$}}
\begin{align}
& \nonumber \mathcal{H}^{(3)}(z, \zb) \Big|_{\log^3(z\zb)}=\frac{(z \zb)^2}{25(z-\zb)^{23}} \left\lbrace R_1(z,\zb) \HPL{001}{z}+R_2(z,\zb)\HPL{011}{z}+R_3(z,\zb)\HPL{101}{z} \right. \\& \nonumber \qquad \left. +R_4(z,\zb) \HPL{01}{z}+R_5(z,\zb) \HPL{11}{z}+\left(R_6(z,\zb) -\frac{\pi ^2 \left(2
   R_3(z,\zb)-R_2(z,\zb)\right)}{6}\right)
   \HPL{1}{z}\right. \\&  \qquad \left. -(z \leftrightarrow \zb)+R_{7}(z,\zb) \right\rbrace \, , \label{correlator}
\end{align}
where $R_i$ are polynomials of degree 30 in $z, \zb$ such that $R_i(z, \zb)=R_i(\zb, z)$ for $i=1,2,3, 7$.\\We compute the flat space limit following the procedure explained in Sec.~\ref{subsec:DDandFlat}: 
\begin{align} \label{cff2l}
&\frac{\text{dDisc}[z \zb (\zb-z)]\mathcal{H}^{(3)}(z^{\circlearrowright}, \zb)]_{\log^3(z \zb)}}{4\pi^2} \xrightarrow{\text{flat space}}2 \pi i \frac{\Gamma(22)}{(2x)^{22}} \log(1-\zb) g^{(3)}( \zb)\, ,\\
\label{g3}
&g^{(3)}( \zb)=\frac{(\zb-1)^6}{3456000 \zb^8} \left\lbrace -60 \left(21 \zb^5-30 \zb^4+10 \zb^3-10 \zb^2+30 \zb-21\right) \HPL{11}{\zb} \right.\\& \nonumber \left. \qquad + 60 \left(10 \zb^2-30 \zb+21\right) \HPL{10}{\zb}-60 \left(21 \zb^2-30 \zb+10\right) \zb^3
   \HPL{01}{\zb} \right.\\& \nonumber \left. \qquad+\left(-2
   \zb^5+1245 \zb^4-1290 \zb^3+1290 \zb^2-60 i \pi  \left(10 \zb^2-30 \zb+21\right)-1245 \zb+2\right)
   \HPL{1}{\zb} \right.\\& \nonumber \left. \qquad-\left(2 \zb^4+15 \zb^3+120 \zb^2-1170 \zb+1260\right) \zb \HPL{0}{\zb}-10 \pi ^2 \left(21 \zb^2-30 \zb+10\right) \zb^3\right.\\& \nonumber \left. \qquad+\left(-1258 \zb^3+871 \zb^2-871 \zb+1258\right) \zb+i \pi  \left(2
   \zb^4+15 \zb^3+120 \zb^2-1170 \zb+1260\right) \zb\right\rbrace \, .
\end{align}
Remember that what we have found is only part of the dDisc, in fact at $\kappa =3$  also $\log^2(z \zb)$ contributes, and since unfortunately we have not access to this information, we can not fully reconstruct the correlator. Given the incompleteness of our answer, we expect that we can not perfectly compare our result  with the amplitude computed on the gravity side. We have  further discussed  these issues is Sec.~\ref{sec:DiscAndCFT}.

\subsection{Two loop SUGRA Amplitude}\label{exp_res_ampl}
The ten dimensional flat space graviton amplitude at two loops has been studied in \cite{Bern:1998ug, Green:2008bf}, however these works concentrate only on  divergent parts, while for our purposes we need to know the full answer, finite part included.
To obtain it, we have to compute the two-loop diagrams (see Fig.~\ref{Fig:boxes}) appearing in Eq.~\eqref{10dAmpl} and to do so we will use the method of differential equations  \cite{Kotikov:1990kg,Kotikov:1991pm,Bern:1993kr,Remiddi:1997ny,Gehrmann:1999as,Henn:2013pwa,%
  Papadopoulos:2014lla,Lee:2014ioa,Ablinger:2015tua,Papadopoulos:2015jft,Liu:2017jxz}.
Let us briefly describe how the method works as it will also be useful
to understand how the discontinuity and the cuts are constructed.
The differential equations are constructed by differentiating the integral of interest with respect to the kinematic invariants, in our case the only dependence is on the ratio $t/s$. 
The integrals appearing after the differentiation are not the starting
ones, but can be reduced to a set of Master Integrals (MI) by integral
by parts identities (IBP) \cite{CHETYRKIN1981159} . So in general one wants to apply the
differentiation on the MI's as all the others can be expressed as a linear combinations of them. 
In the case of interest the integrals contributing are depicted below in Eq.~\eqref{mastersUT}, where the dashed line represents a numerator
constructed from the associated loop momenta and the opposite external
momenta.
The next step in order to solve the system of differential equations is to construct a canonical basis: different results and
algorithms to obtain it can be found in
\cite{Meyer:2017joq,Henn:2020lye,Prausa:2017ltv,Lee:2014ioa,Gituliar:2017vzm}
for the diagrams we are considering. We report our choice for the
planar diagram\footnote{A basis for the non planar one can be found here \cite{Argeri:2014qva}.}, the one we will be focusing on
  \begin{align}\label{mastersUT}
    I_1&=-2 \epsilon  (2 \epsilon -1) (3 \epsilon -2) (3 \epsilon -1) \hspace{2mm} \begin{tikzpicture}[baseline={([yshift=-0.5ex]current
        bounding
        box.center)},scale=0.28,node/.style={draw,shape=circle,fill=black,scale=0.4}]
      \def\xs{12};
\def\ys{10};
   \draw [very thick] (3+3*\xs,2-\ys) ellipse (2cm and 1.5cm);
    \draw [thick] (1+3*\xs,2-\ys) -- (3*\xs,1-\ys);
    \draw [thick] (1+3*\xs,2-\ys) -- (3*\xs,3-\ys);
    \draw [thick] (5+3*\xs,2-\ys) -- (6+3*\xs,1-\ys);
    \draw [thick] (5+3*\xs,2-\ys) -- (6+3*\xs,3-\ys);
    \draw [thick] (1+3*\xs,2-\ys) -- (5+3*\xs,2-\ys);
  \end{tikzpicture} \qquad  I_2=-\frac{s}{t}2 \epsilon  (2 \epsilon -1) (3 \epsilon -2) (3 \epsilon -1) \hspace{2mm} \begin{tikzpicture}[baseline={([yshift=-0.5ex]current
        bounding
        box.center)},scale=0.28,node/.style={draw,shape=circle,fill=black,scale=0.4}]
    \draw [very thick] (2,3) ellipse (1.5cm and 2cm);
    \draw [thick] (2,1) -- (1,0);
    \draw [thick] (2,1) -- (3,0);
    \draw [thick] (2,5) -- (1,6);
    \draw [thick] (2,5) -- (3,6);
    \draw [thick] (2,1) -- (2,5);
  \end{tikzpicture} \nonumber  \\  
          I_3&= \frac{1}{2}\epsilon^2  (2 \epsilon -1) (3 \epsilon -1)
         \hspace{2mm} \begin{tikzpicture}[baseline={([yshift=-0.5ex]current
             bounding box.center)},scale=0.28,node/.style={draw,shape=circle,fill=black,scale=0.4}]
       \def\xs{0};
\def\ys{0};
   \draw [thick] (1+\xs,2-\ys) -- (5+\xs,-\ys);
    \draw[thick] (5+\xs,4-\ys)--(1+\xs,2-\ys);
    \draw [very thick] (5+\xs,2-\ys) ellipse (0.5cm and 2cm);
    \draw [thick] (1+\xs,2-\ys) -- (0+\xs,1-\ys);
    \draw [thick] (1+\xs,2-\ys) -- (0+\xs,3-\ys);
    \draw [thick] (5+\xs,0-\ys) -- (6+\xs,-1-\ys);
    \draw [thick] (5+\xs,4-\ys) -- (6+\xs,5-\ys);
  \end{tikzpicture}
  \qquad \qquad \qquad I_4= \epsilon^2  (2 \epsilon -1)^2
         \hspace{2mm} \begin{tikzpicture}[baseline={([yshift=-0.5ex]current
             bounding box.center)},scale=0.28,node/.style={draw,shape=circle,fill=black,scale=0.4}]
       \def\xs{0};
\def\ys{0};
    \draw [very thick] (1.5,2-\ys) ellipse (1.5cm and 1cm);
   \draw [very thick] (4.5,2-\ys) ellipse (1.5cm and 1cm);
    \draw [thick] (0,2-\ys) -- (-1,1-\ys);
    \draw [thick] (0,2-\ys) -- (-1,3-\ys);
    \draw [thick] (6,2-\ys) -- (7,1-\ys);
    \draw [thick] (6,2-\ys) -- (7,3-\ys);
  \end{tikzpicture} \nonumber \\ I_5&= 3  I_3
                                      +3\epsilon^3  (2 \epsilon -1)
         \hspace{2mm} \begin{tikzpicture}[baseline={([yshift=-0.5ex]current
             bounding box.center)},scale=0.28,node/.style={draw,shape=circle,fill=black,scale=0.4}]
       \def\xs{0};
\def\ys{0};
  \draw [thick] (1+3*\xs,0) -- (5+3*\xs,0);
    \draw[thick] (5+3*\xs,4)--(1+3*\xs,4)--(1+3*\xs,0);
    \draw [very thick] (5+3*\xs,2) ellipse (1cm and 2cm);
    \draw [thick] (1+3*\xs,0) -- (0+3*\xs,-1);
    \draw [thick] (1+3*\xs,4) -- (0+3*\xs,5);
    \draw [thick] (5+3*\xs,0) -- (6+3*\xs,-1);
    \draw [thick] (5+3*\xs,4) -- (6+3*\xs,5);
  \end{tikzpicture} \qquad \qquad \qquad \quad                            
      I_6= -\frac{\epsilon^4  (s+t)}{s}
         \hspace{2mm} \begin{tikzpicture}[baseline={([yshift=-0.5ex]current
             bounding box.center)},scale=0.28,node/.style={draw,shape=circle,fill=black,scale=0.4}]
       \def\xs{0};
\def\ys{0};
   \draw [thick] (1+2*\xs,0) -- (5+2*\xs,0)--(5+2*\xs,4)--(1+2*\xs,4)--(1+2*\xs,0);
    \draw [thick] (5+2*\xs,0) -- (1+2*\xs,4);
    \draw [thick] (1+2*\xs,0) -- (0+2*\xs,-1);
    \draw [thick] (1+2*\xs,4) -- (0+2*\xs,5);
    \draw [thick] (5+2*\xs,0) -- (6+2*\xs,-1);
    \draw [thick] (5+2*\xs,4) -- (6+2*\xs,5);
  \end{tikzpicture}
            \nonumber \\  I_7&= -\frac{t}{s} \epsilon^4  \hspace{2mm} \begin{tikzpicture}[baseline={([yshift=-0.5ex]current
        bounding
        box.center)},scale=0.28,node/.style={draw,shape=circle,fill=black,scale=0.4}]
      \def\xs{0};
\def\ys{0};
     \draw [thick] (0+\xs,0) -- (6+\xs,0)--(6+\xs,3)--(0+\xs,3)--(0+\xs,0);
    \draw [thick] (3+\xs,0) -- (3+\xs,3);
    \draw [thick] (0+\xs,0) -- (-1+\xs,-1);
    \draw [thick] (0+\xs,3) -- (-1+\xs,4);
    \draw [thick] (6+\xs,0) -- (7+\xs,-1);
    \draw [thick] (6+\xs,3) -- (7+\xs,4);
  \end{tikzpicture}
                 \qquad \qquad \qquad \qquad \qquad \quad  I_8=\epsilon^4 \hspace{2mm} \begin{tikzpicture}[baseline={([yshift=-0.5ex]current
        bounding
        box.center)},scale=0.28,node/.style={draw,shape=circle,fill=black,scale=0.4}]
      \def\xs{0};
\def\ys{0};
     \draw [thick] (0+\xs,0) -- (6+\xs,0)--(6+\xs,3)--(0+\xs,3)--(0+\xs,0);
    \draw [thick] (3+\xs,0) -- (3+\xs,3);
    \draw [thick] (0+\xs,0) -- (-1+\xs,-1);
    \draw [thick] (0+\xs,3) -- (-1+\xs,4);
    \draw [thick] (6+\xs,0) -- (7+\xs,-1);
    \draw [thick] (6+\xs,3) -- (7+\xs,4);
    \draw[thick,dashed] (-0.5,1.5)--(3.5,1.5);
  \end{tikzpicture}
 \end{align}
The coefficients are chosen to make them uniform transcendental in $d=4-2\epsilon$ dimensions \cite{Henn:2013pwa}.
From the solution of these differential equation system we were able to find the expressions for the planar and similarly for the non planar double box diagrams. For simplicity, we have chosen the basis integrals and solved the equations in four dimensions. Once the results are know in 4d, it is indeed possible to uplift them to ten dimensions by means of dimensional recurrence relations\footnote{By similar
  reasoning one can prove that the differential equations in Fuchsian
  form is actually $d\pm2$ invariant.} \cite{Tarasov:1996br,Lee:2009dh,Lee:2010wea}.\\ 
We now present the full results for the planar and non planar integrals in 
Fig.~\ref{Fig:boxes}\footnote{As we will see, the $\smallO(\epsilon^{-1})$ does not match exactly the one in \cite{Bern:1998ug} because we have not explicitly  subtracted the one loop pole.}.
\begin{align} 
&I_{db}^{pl}(s, t)=\int \frac{d^{10}l_1}{(2 \pi)^{10}}\, \frac{d^{10}l_2}{(2 \pi)^{10}} \, \frac{1}{l_1^2 (p_1+l_1)^2(p_1+p_2+l_1)^2(l_1+l_2)^2l_2^2(p_4+l_2)^2(p_3+p_4+l_2)^2}\\ \label{resDBPl}&=\frac{(-s)^{3 - 2 \epsilon}}{(4 \pi)^{10}}\left\lbrace -\frac{2}{7!5!}\frac{4s+t}{s \epsilon^2}-\frac{22384 s^3+6247 s^2 t+252 s t^2+63 t^3}{70\,10! s^3\epsilon} +\frac{8 s^6}{5 t^3 (s+t)^3}I^{pl}_{(0)}(s,t)\right\rbrace \, .
\end{align}

The finite part, which is the one relevant for us, is:
\begin{align}\label{Idb_0}
&\nonumber I^{pl}_{(0)}(s,t)=-p_3\left(\sfrac{s}{(s+t)}\right)H_{\text{-}1\text{-}100}+q_3\left(\sfrac{-s}{t}\right)H_{\text{-}100}+P_1(s,t) H_{00}+P_2(s,t)H_{0}+P_3(s,t)\\&  \qquad  +\frac{1}{6}p_3\left(\sfrac{s}{(s+t)}\right)\left(\pi ^2 \left(H_{\text{-}10}-3 H_{\text{-}1\text{-}1}\right)-6 \zeta_3 H_{\text{-}1} \right)+q_3\left(\sfrac{-s}{t}\right)\left(\frac{1}{6} \pi ^2 \left(3 H_{\text{-}1}-H_{0}\right)+\zeta_3\right) \, ,
\end{align}
where $H_{\cdots}\equiv \HPL{\cdots}{\frac{t}{s}}$, $p_3$ and $q_3$ are the functions introduced in Eqs.~\eqref{pPol} and \eqref{qPol}. The $P_i$'s are given by
{\small
\begin{align}
P_1(s,t)&=\frac{2t^2 \left(8820 s^7+40320 s^6 t+72765 s^5 t^2+64575 s^4 t^3+28098 s^3 t^4+4872 s^2 t^5+90 s t^6+10 t^7\right)}{8! \, 5! s^7 (s+t)^2}\\ \nonumber
P_2(s, t)&=-\frac{t^2 \left(52920 s^8+206640 s^7 t+303975 s^6 t^2+202440 s^5 t^3+55111 s^4 t^4+2912 s^3 t^5-29 s^2 t^6-72 s t^7-9 t^8\right)}{10 \, 8! s^9 (s+t)}\\ \nonumber 
P_3(s, t)&=-\frac{t }{15680 \, 10! s^9 (s+t)^2}\left[t (s+t)^2 \left(-51861600 s^7-149838032 s^6 t-135584013 s^5 t^2-24845515 s^4 t^3\right.\right. \\& \nonumber \left. \left.+17602466 s^3 t^4+5332598 s^2 t^5+622251 s t^6+83013 t^7\right)+3920 \pi
   ^2 s^2 \left(17640 s^8+67410 s^7 t+84034 s^6 t^2 \right.\right. \\& \label{polsAmpl}\left. \left.+15834 s^5 t^3-47172 s^4 t^4-37074 s^3 t^5-8544 s^2 t^6-225 s t^7-25 t^8\right)\right]
\end{align}}
For the non planar contribution we obtain
\begin{align}
I_{db}^{np}(s, t)&=\int \frac{d^{10}l_1}{(2 \pi)^{10}}\, \frac{d^{10}l_2}{(2 \pi)^{10}} \, \frac{1}{l_1^2 (p_1+l_1)^2(p_1+p_2+l_1)^2(l_1+l_2)^2l_2^2(p_4+l_2)^2(p_3-l_1-l_2)^2}
\\ \label{resDBNp} &=\frac{(-s)^{3-2\epsilon}}{(4 \pi)^{10}} \left\lbrace \frac{1}{5! \, 6! \epsilon^2}+\frac{917s^2+2t (-s-t)}{3 \, 10!s^2 \epsilon}+\frac{1}{25 \, 9!s^4(s+t)^3} I^{np}_{(0)}(s,t) \right\rbrace \, .
\end{align}
The finite part takes the form
\begin{align} \nonumber
I^{np}_{(0)}(s,t)&=S_1(s,t) \left(H_{\text{-}3}+H_{\text{-}20}-2 
   H_{\text{-}100}\right)+\frac{ (s+t)^3 S_1(s,u)}{t^3}\left(2 H_{\text{-}2\text{-}1}-
   H_{\text{-}1\text{-}2}-
   H_{\text{-}1\text{-}10}\right)\\ & \nonumber +H_{\text{-}10} \left(\frac{1}{60} S_2(s,t)+\frac{i \pi 
   (s+t)^3 S_1(s,u)}{t^3}\right)+H_{-2} \left(\frac{1}{60} S_2(s,t)-\frac{2 i \pi  (s+t)^3
   S_1(s,u)}{t^3}\right)\\ &\nonumber+H_{00} \left(-\frac{(s+t)^3 S_3(s,u)}{60 t^3}-i \pi 
   S_1(s,t)\right)+\frac{1}{60} S_3(s,t) H_{\text{-}1\text{-}1}\\ &\nonumber+H_{-1} \left(\pi ^2 \left(-\frac{(s+t)^3 S_1(s,u)}{2
   t^3}-S_1(s,t)\right)-\frac{1}{60} i \pi  S_3(s,t)+\frac{1}{360} S_4(s,t)\right)\\&\nonumber+H_{0}
   \left(-\frac{(s+t)^3 S_4(s,u)}{360 t^3}+\frac{1}{2} \pi ^2 S_1(s,t)-\frac{1}{60} i \pi 
   S_2(s,t)\right)\\&\nonumber+\frac{1}{120} \pi ^2
   \left(-\frac{(s+t)^3 S_3(s,u)}{t^3}+S_2(s,t)+2100 s^4 (s+t)^3\right)+\zeta (3) S_5(s,t)\\ \label{nonPlanar}&-\frac{1}{2} i
   \pi ^3 S_1(s,t)-\frac{1}{360} i \pi  S_4(s,t)+\frac{S_6(s,t)}{7560} \, ,
\end{align}
where again we have defined the polynomials appearing as:
{\small
\begin{align}
\nonumber
S_1(s,t)&=t^5 \left(15 s^2+10 s t+2 t^2\right)\\ \nonumber
S_2(s,t)&=\frac{(s+t)}{s^4 t^2} \left(-420 s^{12}-2370 s^{11} t-5270 s^{10} t^2-5525 s^9 t^3-2209 s^8 t^4+389 s^7 t^5+171 s^6
   t^6\right. \\& \nonumber \left.+1664 s^5 t^7+3866 s^4 t^8+4500 s^3 t^9+3100 s^2 t^{10}+1200 s t^{11}+200 t^{12}\right) \\ \nonumber
S_3(s,t)&=\frac{(s+t)^8 \left(720 s^8-1280 s^7 t+249 s^6 t^2+78 s^5 t^3-366 s^4 t^4+400 s^3 t^5-300 s^2 t^6+200 s
   t^7-200 t^8\right)}{s^4 t^5} \\ \nonumber
S_4(s,t)&=\frac{(s+t)^2}{s^3 t^4} \left(-4320 s^{12}-16080 s^{11} t-4974 s^{10} t^2+61465 s^9 t^3+120276 s^8 t^4+88029 s^7
   t^5\right. \\& \nonumber \left.+15996 s^6 t^6-7880 s^5 t^7+1296 s^4 t^8+9800 s^3 t^9+10000 s^2 t^{10}+5400 s t^{11}+1200
   t^{12}\right) \\ \nonumber
S_5(s,t)&=3 \left(40 s^7+110 s^6 t+84 s^5 t^2+15 s^2 t^5+10 s t^6+2 t^7\right) \\ \nonumber
S_6(s,t)&= \frac{1}{s^2 t^3}\left(45360 s^{12}+236880 s^{11} t+280917 s^{10} t^2+24169796 s^9 t^3+72161380 s^8 t^4+71728624 s^7
   t^5\right. \\& \nonumber \left.+23275934 s^6 t^6-639392 s^5 t^7-369323 s^4 t^8-278250 s^3 t^9-194250 s^2 t^{10}-75600 s
   t^{11}-12600 t^{12}\right)
\end{align}}

Let us briefly comment on the amplitude divergences: as we can see from Eq.~\eqref{resDBPl} and \eqref{resDBNp}, at two loops  the combination  $(I_{db}^{pl}(s,t)+I_{db}^{np}(s,t)+ t \leftrightarrow u)$, appearing in the amplitude, cancels the divergence at order $\epsilon^{-2}$ , thus making the discontinuity convergent. At higher loops, instead, we do expect a divergence  of order $1/\epsilon^2$ and consequently  a divergent discontinuity of $\mathcal{A}_{10}^{sugra}$. So in principle for $\kappa>3$, if we wanted to compare dDisc with the  discontinuity of the amplitude, it would be necessary to renormalize the latter subtracting the lower loops poles. At the same time, it exists an object derived from the amplitude, which is always finite in ten dimensions, the iterated $s$-channel cut mentioned before, this can be naturally compared to the CFT without the need of any renormalization procedure.
\subsection{Match Amplitude and CFT at 2 loops}\label{ampl_cft_comp}
Here we report the results for the discontinuity of $\mathcal{A}_{10}^{sugra}$ at two loops and its comparison with the flat space limit of dDisc$\mathcal{H}^{(3)}|_{\log^3(z \zb)}$ as described in Sec.~\ref{sec:DiscAndCFT}.
 
The first important remark is that the quantity computed from the CFT correlator should not be compared with the discontinuity of the entire amplitude, but rather with only one channel and restricting to the planar contribution. In particular, in our convention, we have to consider  disc$_{x>1} F_{3, pl}^t(x)$, where $F_{3, pl}^t(x)$ contains the contributions from $t^2(I_{db}^{pl}(t, s)+ s\leftrightarrow u)$.  Before performing the discontinuity, we have defined the amplitude in the correct physical region through the appropriate analytic continuations. Finally to connect with the CFT results, we have  expressed everything as a function of $\zb$, through the identification $\zb=\frac{1}{x}$.
We collect the results of this comparison in the table below, where we have defined $\mathcal{Q}_{CFT} \equiv \frac{i \pi  (\zb-1)^{11}}{10 \zb^8}\log(1-\zb)g^{(3)}(\zb)$ and $\mathcal{Q}_{Ampl}\equiv  \frac{i \pi  (\zb-1)^{11}}{10 \zb^8}\text{disc}F_{3, pl}^t(\zb)$.\\
\newline
\begin{adjustbox}{width=1\textwidth}
\begin{tabular}{c||c|c|c|c|c|c|c|c|c|c}
 & $H_{111}$ &$H_{110}$&$H_{011}$ &$H_{101}$&  $H_{001}$ &
$H_{11}$ &$H_{10}$ &$H_{01}$ &$H_{1}$ & $H_{0}$ \\
 \hline 
$\mathbf{\mathcal{Q}_{CFT}}$& $-3 \left(\tilde{p}_3+p_3\right)$ & $-2 p_3$ & $-2 \tilde{p}_3$ & $-\tilde{p}_3-p_3$ & $0$ & $2 \left(\tilde{q}_3+q_3 +i \pi  p_3\right)$ &$ \tilde{q}_3 $& $\tilde{q}_3 $& $-j_1-i \pi  \tilde{q}_3-\frac{\pi^2}{6} \tilde{p}_3 $&$ 0$ \\[0.5em]
\hline
$\mathbf{\mathcal{Q}_{Ampl}}$&$  -\tilde{p}_3-p_3$ &$ -p_3$ &$ -\tilde{p}_3 $&$ -\tilde{p}_3$ &$ -\tilde{p}_3$ &$ \tilde{q}_3+q_3+i \pi  p_3$ &$ \tilde{q}_3$ &$ q_3 $&$ \tilde{P}_1+P_1-i \pi  \tilde{q}_3+\frac{\pi ^2
   p_3}{6}$ &$ \tilde{P}_1$
\end{tabular}
\end{adjustbox}\\
\newline
Here $H_{\cdots}\equiv H_{\cdots}(\zb)$ is the Harmonic polylogarithm,   $p$ and $q$ are the polynomials  in Eqs. ~\eqref{pPol} and \eqref{qPol}, while $j_1$ and $P_1$ are respectively introduced in Eq.~\eqref{j1} and \eqref{polsAmpl}. For convenience we have defined
$p_3(1-\zb)\equiv p_3$ and $p_3\left(\frac{\zb-1}{\zb}\right)\equiv
\tilde{p}_3$ and analogously for all the other polynomials.\\
As already said, we notice that the two contributions do not match exactly: some terms do not appear or appear with different coefficients on the two sides. Interestingly if one extracts the logarithmic singularities from the HPL's, as explained in Appendix \ref{sec:HarmonicPolylog}, and ignoring weight one and zero terms, all the functions multiplying $\log^n(1-\zb)$ in the CFT actually coincide with the ones appearing in the amplitude, but with a factor $n$ of difference (as can be already deduced from the table above). This fact is another indication of  why $c_{dc}$ reproduces $g^{(3)}(\zb)$ upon the log-factorization. {In particular the factor of three of difference in the leading $\log (1-\zb)$ term is exactly the one necessary to have their perfect agreement.}

The contributions to the discontinuity coming from the other channels in $\mathcal{A}_{10}^{sugra}$ at order $c^{-3}$ can be partially recovered considering crossing symmetric versions \cite{Dolan:2004iy} of $\mathcal{H}^{(3)}|_{\log^3(z \zb)}$, to be precise
\begin{gather}
\mathcal{H}^{(3)}(u,v)=\frac{u^2}{v^2} \mathcal{H}^{(3)}(v,u) \leftrightarrow s\text{ channel} \, , \\
\mathcal{H}^{(3)}(u,v)=\frac{u^2}{v^2} \mathcal{H}^{(3)}\left(\frac{1}{v},\frac{u}{v}\right) \leftrightarrow u\text{ channel} \, .
\end{gather} 
Also in these two other channels the match is not perfect and we encounter the same type of discrepancies found before. 

We can now pass to analyse the $c_1$ and $c_2$ type of discontinuities, they can be computed exploiting the same differential equation method introduced before. Cut diagrams do indeed satisfy the same system of equations \cite{Bosma:2017hrk,Frellesvig:2017aai} but now with
different boundary conditions and where only a sub-set of topologies
contribute. To be more precise, the diagrams contributing to each cut will be the ones where the onshell propagators are present and in the case at hand we find
\begin{align}
&\nonumber
  c_1: \qquad \begin{tikzpicture}[baseline={([yshift=-1.1ex]current bounding box.center)},scale=0.2,node/.style={draw,shape=circle,fill=black,scale=0.4}]
   \draw [thick] (0,0) -- (6,0)--(6,3)--(0,3)--(0,0);
    \draw [thick] (3,0) -- (3,3);
    \draw [thick] (0,0) -- (-1,-1);
    \draw [thick] (0,3) -- (-1,4);
    \draw [thick] (6,0) -- (7,-1);
    \draw [thick] (6,3) -- (7,4);
    \draw [dashed,thick,red] (1.5,4)--(1.5,-1);
  \end{tikzpicture}\rightarrow \begin{tikzpicture}[baseline={([yshift=-1.1ex]current bounding box.center)},scale=0.2,node/.style={draw,shape=circle,fill=black,scale=0.4}]
   \draw [thick] (0,0) -- (6,0)--(6,3)--(0,3)--(0,0);
    \draw [thick] (3,0) -- (3,3);
    \draw [thick] (0,0) -- (-1,-1);
    \draw [thick] (0,3) -- (-1,4);
    \draw [thick] (6,0) -- (7,-1);
    \draw [thick] (6,3) -- (7,4);
  \end{tikzpicture}+\begin{tikzpicture}[baseline={([yshift=-1.1ex]current bounding box.center)},scale=0.2,node/.style={draw,shape=circle,fill=black,scale=0.4}]
  	\def\xs{0};
     \draw [thick] (1+3*\xs,0) -- (5+3*\xs,0);
    \draw[thick] (5+3*\xs,4)--(1+3*\xs,4)--(1+3*\xs,0);
    \draw [ thick] (5+3*\xs,2) ellipse (1cm and 2cm);
    \draw [thick] (1+3*\xs,0) -- (0+3*\xs,-1);
    \draw [thick] (1+3*\xs,4) -- (0+3*\xs,5);
    \draw [thick] (5+3*\xs,0) -- (6+3*\xs,-1);
    \draw [thick] (5+3*\xs,4) -- (6+3*\xs,5);
  \end{tikzpicture}+\begin{tikzpicture}[baseline={([yshift=-1.1ex]current bounding box.center)},scale=0.2,node/.style={draw,shape=circle,fill=black,scale=0.4}]
  \def\ys{0};
     \draw [ thick] (1.5-1,2-\ys) ellipse (1.5cm and 1cm);
   \draw [ thick] (4.5-1,2-\ys) ellipse (1.5cm and 1cm);
    \draw [thick] (0-1,2-\ys) -- (-1-1,1-\ys);
    \draw [thick] (0-1,2-\ys) -- (-1-1,3-\ys);
    \draw [thick] (6-1,2-\ys) -- (7-1,1-\ys);
    \draw [thick] (6-1,2-\ys) -- (7-1,3-\ys);
  \end{tikzpicture}+\begin{tikzpicture}[baseline={([yshift=-1.1ex]current bounding box.center)},scale=0.2,node/.style={draw,shape=circle,fill=black,scale=0.4}]
  \def\ys{0};
  \def\xs{0};
    \draw [thick] (1+\xs,2-\ys) -- (5+\xs,-\ys);
    \draw[thick] (5+\xs,4-\ys)--(1+\xs,2-\ys);
    \draw [ thick] (5+\xs,2-\ys) ellipse (0.5cm and 2cm);
    \draw [thick] (1+\xs,2-\ys) -- (0+\xs,1-\ys);
    \draw [thick] (1+\xs,2-\ys) -- (0+\xs,3-\ys);
    \draw [thick] (5+\xs,0-\ys) -- (6+\xs,-1-\ys);
    \draw [thick] (5+\xs,4-\ys) -- (6+\xs,5-\ys);
  \end{tikzpicture} \, , \\&
\nonumber
 c_2: \qquad \begin{tikzpicture}[baseline={([yshift=-1.1ex]current bounding box.center)},scale=0.2,node/.style={draw,shape=circle,fill=black,scale=0.4}]
   \draw [thick] (0,0) -- (6,0)--(6,3)--(0,3)--(0,0);
    \draw [thick] (3,0) -- (3,3);
    \draw [thick] (0,0) -- (-1,-1);
    \draw [thick] (0,3) -- (-1,4);
    \draw [thick] (6,0) -- (7,-1);
    \draw [thick] (6,3) -- (7,4);
    \draw [dashed,thick,red] (1.5,4)--(4.5,-1);
  \end{tikzpicture}\rightarrow \begin{tikzpicture}[baseline={([yshift=-1.1ex]current bounding box.center)},scale=0.2,node/.style={draw,shape=circle,fill=black,scale=0.4}]
   \draw [thick] (0,0) -- (6,0)--(6,3)--(0,3)--(0,0);
    \draw [thick] (3,0) -- (3,3);
    \draw [thick] (0,0) -- (-1,-1);
    \draw [thick] (0,3) -- (-1,4);
    \draw [thick] (6,0) -- (7,-1);
    \draw [thick] (6,3) -- (7,4);
  \end{tikzpicture}+\begin{tikzpicture}[baseline={([yshift=-1.1ex]current bounding box.center)},scale=0.2,node/.style={draw,shape=circle,fill=black,scale=0.4}]
  	\def\xs{0};
      \draw [thick] (1+2*\xs,0) -- (5+2*\xs,0)--(5+2*\xs,4)--(1+2*\xs,4)--(1+2*\xs,0);
    \draw [thick] (5+2*\xs,4) -- (1+2*\xs,0);
    \draw [thick] (1+2*\xs,0) -- (0+2*\xs,-1);
    \draw [thick] (1+2*\xs,4) -- (0+2*\xs,5);
    \draw [thick] (5+2*\xs,0) -- (6+2*\xs,-1);
    \draw [thick] (5+2*\xs,4) -- (6+2*\xs,5);
  \end{tikzpicture}+\begin{tikzpicture}[baseline={([yshift=-1.1ex]current bounding box.center)},scale=0.2,node/.style={draw,shape=circle,fill=black,scale=0.4}]
  \def\ys{0};
  \def\xs{0};
   \draw [thick] (3+3*\xs,2-\ys) ellipse (2cm and 1.5cm);
    \draw [thick] (1+3*\xs,2-\ys) -- (3*\xs,1-\ys);
    \draw [thick] (1+3*\xs,2-\ys) -- (3*\xs,3-\ys);
    \draw [thick] (5+3*\xs,2-\ys) -- (6+3*\xs,1-\ys);
    \draw [thick] (5+3*\xs,2-\ys) -- (6+3*\xs,3-\ys);
    \draw [thick] (1+3*\xs,2-\ys) -- (5+3*\xs,2-\ys);
  \end{tikzpicture}  \, .
\end{align}
Each solution is fixed up to a boundary condition that
unfortunately we were not able to fix separately, 
 we could express one condition in term of the other by requiring that
\begin{align}\label{amp2}
&\text{disc} _s I_{db}=2 \pi i \left(\mathcal{A}\big|_{c_1}+\mathcal{A}\big|_{c_2}\right) \, .
\end{align}
However, since in the CFT we expect that higher $\log(1-\zb)$ powers are fixed by double trace contributions, we can use this insight to partially fix these boundary conditions. We can compare again dDisc$\mathcal{H}^{(3)}|_{\log^3( z \zb)}$ with disc$\mathcal{A}_{10}^{sugra}$, but now we can distinguish and analyse separately $c_1$ and $c_2$. The table below contains the result of this comparison, we report the weight three and two functions appearing in Eq.~\eqref{cff2l} and in $c_1$ and $c_2$ respectively with the corresponding polynomial coefficients properly normalized.
\vspace{5mm}\\
\begin{adjustbox}{width=1\textwidth}
\small
\begin{tabular}{c||c|c|c|c|c|c|c|c|c|c}
& $H_{111}$ & $H_{110}$ & $H_{011}$& $H_{101}$& $H_{001}$& $H_{100}$& $H_{11}$& $H_{10}$& $H_{01}$& $H_{00}$
\\[0.5em]
\hline
\textbf{CFT} & $-3 \left(p_3+\tilde{p}_3\right)$& $-2p_3$ & $-2 \tilde{p}_3$ & $-(p_3+\tilde{p}_3)$& & &$2(q_3+\tilde{q}_3)$ &$\tilde{q}_3$& $\tilde{q}_3$&  \\[0.5em]
\hline 
$\mathbf{c_1}$& $- \left(p_3+\tilde{p}_3\right)$ & $-p_3$ & $-\tilde{p}_3$ & $-p_3$ & &$-p_3$ & $q_3+\tilde{q}_3$ & $\tilde{q}_3$&$\tilde{q}_3$ &$\tilde{q}_3$
\\[0.5em]
\hline
$\mathbf{c_2}$& & & & $p_3-\tilde{p}_3$&$-\tilde{p}_3$ &$p_3$& & & $q_3-\tilde{q}_3$ & $-\tilde{q}_3$
\\[0.5em]
\end{tabular}
\end{adjustbox}
\vspace{5mm} 
\newline
where again all $H$ are $H_{\cdots}(\zb)$.\\
The two cuts present lots of similarities in the polynomial
structure but a different singular behaviour. The similarities of the
polynomial structure are expected as they come from the same starting
integral and it partially  explains why, when they mix in
the full discontinuity, give rise to something which resembles the CFT
computation but with all the discrepancies we observe. Even after
having separated the two contributions, $c_1$ still does not 
reproduce the first row of this table, we believe that these differences  derive from part of the $\log^2 U$ term in the correlator.

The last object we have introduced in Sec.~\ref{sec:DiscAndCFT} is the \textit{double cut}. Analogously to what we have done before, this can be constructed from the differential equations where now only one
subtopology contributes,
\begin{equation}
\label{dcpicture}
  c_{dc}: \qquad \begin{tikzpicture}[baseline={([yshift=-1.1ex]current bounding box.center)},scale=0.25,node/.style={draw,shape=circle,fill=black,scale=0.4}]
   \draw [thick] (0,0) -- (6,0)--(6,3)--(0,3)--(0,0);
    \draw [thick] (3,0) -- (3,3);
    \draw [thick] (0,0) -- (-1,-1);
    \draw [thick] (0,3) -- (-1,4);
    \draw [thick] (6,0) -- (7,-1);
    \draw [thick] (6,3) -- (7,4);
    \draw [dashed,thick,red] (1.5,4)--(1.5,-1);
    \draw [dashed,thick,red] (4.5,4)--(4.5,-1);
  \end{tikzpicture}\rightarrow \begin{tikzpicture}[baseline={([yshift=-1.1ex]current bounding box.center)},scale=0.25,node/.style={draw,shape=circle,fill=black,scale=0.4}]
   \draw [thick] (0,0) -- (6,0)--(6,3)--(0,3)--(0,0);
    \draw [thick] (3,0) -- (3,3);
    \draw [thick] (0,0) -- (-1,-1);
    \draw [thick] (0,3) -- (-1,4);
    \draw [thick] (6,0) -- (7,-1);
    \draw [thick] (6,3) -- (7,4);
  \end{tikzpicture}+\begin{tikzpicture}[baseline={([yshift=-1.1ex]current bounding box.center)},scale=0.25,node/.style={draw,shape=circle,fill=black,scale=0.4}]
  \def\ys{0};
     \draw [thick] (1.5-1,2-\ys) ellipse (1.5cm and 1cm);
   \draw [thick] (4.5-1,2-\ys) ellipse (1.5cm and 1cm);
    \draw [thick] (0-1,2-\ys) -- (-1-1,1-\ys);
    \draw [thick] (0-1,2-\ys) -- (-1-1,3-\ys);
    \draw [thick] (6-1,2-\ys) -- (7-1,1-\ys);
    \draw [thick] (6-1,2-\ys) -- (7-1,3-\ys);
  \end{tikzpicture}\,.
\end{equation}
With a suitable choice of boundary conditions we can fix $c_{dc}$ in
such a way that it matches Eq.~\eqref{cff2l} up to a $\log(1-\zb)$ factorization as in Eq.~\eqref{doublecutandg3}. 
Let us comment on the advantages of computing  this iterated cut directly from the differential equations: it is computationally easier because we can reuse the results from the previous computations and most importantly in this way we avoid  the need to specify any $i \epsilon$ prescription, that are in general not well defined in the case of multiple discontinuities. 

{To conclude, let us spend a few words on the possible contributions of the non planar double box diagram. As can be derived from Eq.~\eqref{nonPlanar}, the integral can contain only terms of transcendental weight two or lower, to be precise the subset of HPLs appearing in the $t$-channel is made by $\{H_{01}, \, H_{10}, \, H_{00}, \, H_{1}, \, H_{0}\}$.  The reasons why we have decided not to  include it  in the previous discussion are one practical and one more conceptual. The first one comes from the direct computation of the discontinuity that gives polynomials with a structure completely different from the ones we have been considering. The second one relies on the fact that we believe that the discontinuity associated to the non planar box should be reproduced by part of the $\log^2 U$ piece of $\mathcal{H}^{(3)}$ that  unfortunately we are not able to compute. To do so it would be indeed necessary to completely solve the mixing problem up to order $c^{-2}$ and to find a way to include triple trace operators and unfortunately neither of these problems have been tackled yet.}
\subsection{Two-loop Mellin amplitude} \label{Appendix:Mell}
As anticipated in Sec.~\ref{sec:Mell2l}, for the Mellin amplitude associated to $\mathcal{H}^{(3)}|_{\log^3 (z \zb)}$, which we remind behaves like as $\sim \log^3 U \log^2 V$ in the double expansion around small $U$ and $V$, one can consider the following ansatz: 
\begin{align}
\label{mell2loopExpr}
\mathcal{\widetilde{M}}^{(3)}_{\log} (s, t)=\sum_{m,n=2} \frac{c_{mn}}{(s-2m)^2 (t-2n)}+ \text{cyclic permutation} \, .
\end{align}
Plugging this expression, without considering the permutations, in  Eq.~\eqref{MellTranform} and comparing the resulting $\log^3 U \log^2V$ terms with the corresponding ones in the correlator of Eq.~\eqref{correlator}, one can fix the $c_{mn}^{(3)}$:
\begin{align}\label{cmn}
c_{mn}^{(3)}=\frac{r^{(12)}(m,n) H_{m+n-1}}{(n-1)_7}+\sum_{i=0}^6\left(\frac{r_{1,i}^{(8)}(m)H_{m-i}}{n+i-1}+\frac{r_{2,i}^{(8)}(m)}{m+n-1-i} \right) \, .
\end{align}
The $r^{(j)}$ are polynomials of degree $j$ in the corresponding variables, see the ancillary Mathematica file for their explicit expressions. 

Now the flat space limit can be taken allowing $s, \, t $ to go to infinity, however, given the form of Eq.~\eqref{mell2loopExpr}, this is not obvious and needs to be taken carefully. Retracing the same steps of \cite{Alday:2018kkw}, let us start by focusing on one channel of our Mellin amplitude and consider
\begin{align}
\mathcal{\widetilde{M}}^{(3)}_{\log,1ch}=\sum_{m,n=2}^{\infty}\frac{c_{mn}^{(3)}}{(s-2m)^2(t-2n)} \, .
\end{align} 
First of all notice that this series is divergent. To  regularize it, we take as many derivatives in $s$ and $t$ as the ones necessary to make it convergent\footnote{We have verified that keeping the total number of derivatives fixed, reshuffling between $s$ and $t$ derivatives determines only the appearance of a factor of $\frac{s}{t}$ to some power and a change in the  polynomial term.}. Then, the flat space limit is implemented by taking $s,t \to \infty$ with $m$ and $n$ of the same order. In this regime, the sums can be replaced by integrals and eventually we are left with
\begin{align}
\label{eq:intMell}
 \partial_s^2 \partial_t^3\mathcal{\widetilde{M}}^{(3)}_{\log,1ch} \simeq \int_0^{\infty} dm \, dn \, \frac{-10 ! m^{11}}{20 n^3 (m+n)^3} \,  \frac{-p_3(\sfrac{m}{m+n})\log\left(1+\frac{n}{m} \right)+q_3(-\sfrac{m}{n})}{(s-2m)^4(t-2n)^4} \, ,
\end{align}
where the polynomials $p_3$ and $q_3$ are defined in Eq.~\eqref{p3} and \eqref{q3}\footnote{The same result was recently obtained in \cite{Aprile:2020luw} in a completely different set-up.}.
Notice the similarity with the amplitude (the same permutations are involved) and in particular with the form of the planar double box integral in Eq.~\eqref{Idb_0}.\\In analogy with the one-loop example, one would expect that, when integrated in $m$ and $n$, Eq.~\eqref{eq:intMell} gives derivatives of the two-loop amplitude in the corresponding channel\footnote{In this context, we has to consider the amplitude as in Eq.~\eqref{10dAmpl} without the $\hat{K}$ factor, as we have already accounted for it in $\Theta_4^{flat}$. Moreover, notice that this number of derivatives kills the divergent parts.} or at least part of it. However this seems not to be the case.\\ Let us start considering the integral of Eq.~\eqref{eq:intMell}, this reads
\begin{align} \label{intMellDer}
 \partial_s^2 \partial_t^3\mathcal{\widetilde{M}}^{(3)}_{\log,1ch}=\frac{1}{s}\left( f_1(\sfrac{t}{s})\left(H_{-100}+\frac{\pi^2}{2}H_{-1}\right)+f_2(\sfrac{t}{s})\left(H_{00}+\frac{\pi^2}{2}\right)+f_3(\sfrac{t}{s})H_{0}+f_4(\sfrac{t}{s}) \right)
\end{align}	 
with $H_{\ldots}\equiv \HPL{\ldots}{\frac{t}{s}}$ . If we compare the previous expression with the derivatives of the amplitude, we observe that:
\begin{enumerate}[i)]
\itemsep-1em
\item The maximum degree of transcendentality of the functions in Eq.~\eqref{intMellDer} is one degree less than the one in the amplitude, where even after taking derivatives, they appear HPL's of weight four and lower.\\
\item If one imagines to integrate back the derivatives in $s$ and $t$, then $\mathcal{\widetilde{M}}^{(3)}_{\log, 1ch} \sim s^4$, which does not match again the amplitude ($\sim s^5$). Moreover,  this behaviour seems to be in contrast with the general prescription in \cite{Penedones:2019tng}: here the authors find a relation between the way in which the correlator diverges in the bulk point limit and the polynomial growth of the Mellin amplitude. According to their prediction a singularity $\sim (z-\zb)^{-23}$, which is the one we are considering, should produce a Mellin amplitude that goes\footnote{If one applies formula (217) of \cite{Penedones:2019tng} with $\Delta=2$, one find $s^9$, however we need to remember that in $\mathcal{\widetilde{M}}$ we have stripped out a factor $s^4$, so the two results are consistent.} as $s^5$, in accordance with the flat space amplitude result we have.\\
\item If one ignores what said before and nonetheless tries to enforce a comparison with the amplitude, one soon realizes that, in order to at least reproduce the structure of the polynomials $f_i$, it is necessary to take one $s$ derivative more of the amplitude, i.e. to consider $\partial_s^3 \partial_t^3 \left( s^2 I_{db}^{pl}(s, t)\right)$. This indeed gives
\begin{align} \nonumber 
&\partial_s^ 3 \partial_t^3  \left( s^2 I_{db}^{pl}(s, t)\right)=\frac{16}{14175\, s} \left(f_1(\sfrac{t}{s}) \left(H_{-1-100}  -\frac{\pi ^2}{6} H_{-10} +\frac{\pi ^2}{2} H_{-1-1}+ \zeta_3 H_{-1}\right)\right. \\ & \nonumber \left. +f_2(\sfrac{t}{s})\left(H_{-100}+\frac{ \pi ^2}{2} H_{-1}-\frac{\pi ^2}{6}  H_{0}+\zeta_3\right)+(\tilde{f}_3(\sfrac{t}{s})-\tilde{\tilde{f}}_3(\sfrac{t}{s}))H_{00} +\tilde{f}_4(\sfrac{t}{s})H_0  \right. \\ &   \left.-\tilde{\tilde{f}}_4(\sfrac{t}{s}))H_{0}  +f_5(\sfrac{t}{s}) \pi^2+ f_6(\sfrac{t}{s})\right) \label{derAmpl} \, .
\end{align}
\end{enumerate}
These observations motivated the study of the maximal cut in Eq.~\eqref{2loopsMdc}. Then reinterpreting the problem from this perspective and in light of our discussion in Sec.~\ref{sec:Mell2l}, we should  have somehow expected not to reproduce the full amplitude starting from \eqref{eq:intMell}, in fact there is not a dispersion relation representation constructed from such multiple discontinuities. Moreover, given the fact that we are taking only a double integral, it seems more  reasonable why we should take an additional derivative and why we do not get functions with the right transcendentality. For all these reasons we have decided to try to compare Eq.~\eqref{intMellDer} with our results for $c_1$ and $c_2$, looking for some similarities and connections. So let us consider

\begin{align*}
\partial_s\left(\partial_s^2 \partial_t^3 \begin{tikzpicture}[baseline={([yshift=-1.1ex]current bounding box.center)},scale=0.25,node/.style={draw,shape=circle,fill=black,scale=0.4}]
   \draw [thick] (0,0) -- (6,0)--(6,3)--(0,3)--(0,0);
    \draw [thick] (3,0) -- (3,3);
    \draw [thick] (0,0) -- (-1,-1);
    \draw [thick] (0,3) -- (-1,4);
    \draw [thick] (6,0) -- (7,-1);
    \draw [thick] (6,3) -- (7,4);
    \draw [dashed,thick,red] (1.5,4)--(1.5,-1);
  \end{tikzpicture} \right)_{\pi^0 \text{ terms}}&=\frac{16}{14175 \, s}\left( f_1(\sfrac{t}{s}) H_{-100}+f_2(\sfrac{t}{s}) H_{00}+ \tilde{f}_3(\sfrac{t}{s}) H_0+\tilde{f}_4(\sfrac{t}{s})  \right)\, ,\\
\partial_s\left(\partial_s^2 \partial_t^3 \begin{tikzpicture}[baseline={([yshift=-1.1ex]current bounding box.center)},scale=0.25,node/.style={draw,shape=circle,fill=black,scale=0.4}]
   \draw [thick] (0,0) -- (6,0)--(6,3)--(0,3)--(0,0);
    \draw [thick] (3,0) -- (3,3);
    \draw [thick] (0,0) -- (-1,-1);
    \draw [thick] (0,3) -- (-1,4);
    \draw [thick] (6,0) -- (7,-1);
    \draw [thick] (6,3) -- (7,4);
    \draw [dashed,thick,red] (1.5,4)--(4.5,-1);
  \end{tikzpicture}\right)_{\pi^0 \text{ terms}}&=\frac{-16}{14175 \, s}\left( f_1(\sfrac{t}{s})( H_{-100}-H_{-1-10})+f_2(\sfrac{t}{s}) (H_{00}-H_{-10})\right. \\ & \left.+ \tilde{\tilde{f}}_3(\sfrac{t}{s}) H_0+\tilde{\tilde{f}}_4(\sfrac{t}{s})  \right) \, .
 \end{align*}
Comparing them with Eq.~\eqref{intMellDer}, we notice that the highest transcendentality pieces here are correctly reproduced by $c_1$. The results for $c_2$ are very similar, but this is inevitable since they come from the same terms in the amplitude. The relevant fact is  instead  that if we sum up the two contributions the characteristic structure of Eq.~\eqref{intMellDer} is completely lost, thus signalling again the need to distinguish the two types of cuts. The qualitative idea motivating the comparison of the Mellin result and the additional derivative of the discontinuities is that starting from a double pole in $s$ and a simple one in $t$, performing a double integral is not sufficient and we are left with a $s$ pole still to be solved. However these are, as said, qualitative observations based on the example at hand and we were not able to get to a better and more rigorous  explanation.

Let us conclude by a brief comment on the lower $\log U , \, \log V$-powers coming from the residue integral defining the Mellin transform, once we plug in the expression for $\mathcal{\widetilde{M}}^{(3)}_{\log,1ch}$. As anticipated in Sec.~\ref{sec:Mell2l}, 
\begin{align}
\mathcal{H}^{(3)}(U, V)=\sum_{m,n=2}^{\infty}&\frac{(n-1)_{m+1}^2 c_{mn}}{24 \Gamma(m-1)^2}U^m V^{n-2} \log^3 U \log^2 V+\\ \nonumber & \frac{
   (n-1)_{m+1}^2 c_{mn}}{4 \Gamma (m-1)^2}U^m V^{n-2} \left(H_{m+n-1}-H_{m-2}\right) \log^2 U \log^2 V+ \cdots \, .
\end{align}
The second term is the one we expect to derive from simple poles in $s$ and $t$, so we argue that the unknown parts of $\mathcal{H}^{(3)}$ should have a Mellin representation including
\begin{align}
\label{eq:MellinSinglePoles}
\sum_{m,n} \frac{6 c_{mn}^{(3)}(H_{m+n-1}-H_{m-2})+b_{mn}^{(3)}}{(s-2m)(t-2n)} \, ,
\end{align}
where $b_{mn}^{(3)}$ are some coefficients that should in principle be fixed by $\log^2 U \log^2 V$ 
terms in the correlator. Even if we do not have access to them, we can still say that, in order to mix with the contributions coming from the $c_{mn}^{(3)}$, the $b_{mn}^{(3)}$ should show something in common and  present a very constrained polynomial structure. For these reasons, we thought that it was worthwhile to study the flat space limit of \eqref{eq:MellinSinglePoles} with $b_{mn}^{(3)}=0$. After having regularized the sum, this reads
\begin{align} \label{singlePoles}
 \partial_s^3 \partial_t^3 \iint dm \, dn \,  \frac{7! \, 6^3 m^{11} \left(2 p_3\left(\sfrac{m}{m+n}\right) H_{-1-1}-q_3\left(\sfrac{-m}{n}\right)
 H_{-1}\right)}{n^3 (m+n)^3 (s-2m) (t-2n)}\,.
\end{align}
Consistently to what we discussed before, we compared the integrand, without the $s, \, t$ poles,  to
\begin{align}
\frac{1}{\pi^2}\begin{tikzpicture}[baseline={([yshift=-1.1ex]current bounding box.center)},scale=0.25,node/.style={draw,shape=circle,fill=black,scale=0.4}]
   \draw [thick] (0,0) -- (6,0)--(6,3)--(0,3)--(0,0);
    \draw [thick] (3,0) -- (3,3);
    \draw [thick] (0,0) -- (-1,-1);
    \draw [thick] (0,3) -- (-1,4);
    \draw [thick] (6,0) -- (7,-1);
    \draw [thick] (6,3) -- (7,4);
    \draw [dashed,thick,red] (1.5,4)--(1.5,-1);
    \draw [dashed,thick,red] (7,1.5)--(-1,1.5);
  \end{tikzpicture} =\frac{32 s^{11} \left(p_3\left(\sfrac{s}{s+t}\right) H_{-1-1}-q_3\left(\sfrac{-s}{t}\right)
   H_{-1}-P_1(s,t)\right)}{5 t^3 (s+t)^3} \, .
\end{align}
Again, notice the mismatch in the multiplicative coefficients of the highest transcendentality pieces. The reason behind it should again be found in the fact that we do not know triple trace operators contributions and we should in principle consider  the $b_{mn}^{(3)}$. Similarly, if we integrate Eq.~\eqref{singlePoles}, we obtain part of the amplitude (or part of $c_1$ if we take its discontinuity) with the right weights, but with some mismatches. Nevertheless, all these similarities tell us that, at least in the flat space limit, the form of the full correlator is very constrained and that the unknown terms should mix with the studied ones in a very precise way and with a similar polynomial structure.
\section{Polynomial Structure } \label{appendix:polStructure}
In Sec.~\ref{sec:FSLeadingLogs} we have reported the expressions for the polynomials of the highest and next to the highest transcendental functions, namely $p_{\kappa}$ and $q_{\kappa}$ in Eqs.~\eqref{pPol} and \eqref{qPol}. The same contributions can be extracted from the iterated $s$-channel cut computations in Sec.~\ref{sec_higher_loops}, exploiting the iterative method we have developed. 
Let us now explain how we can extract the highest weight polynomials:
\begin{itemize}
  \item the integrand is splitted into two regions, each one has an associated kernel $\mathcal{K}^{\pm}(v,x)$ where we have defined the adimensional parameter $x=t/s$. After each integration $x$ is mapped to $1-v$.
  \item we are interested in finding the poles, as those integrated
    will give the highest transcendental function;
  \item at each loop order only one of the two $\mathcal{K}^{\pm}(v,x)$
  functions will contribute with a pole:  the $-$ with a pole in $v=0$ for the even loops and the $+$ with a pole in $v=1$ for the odd ones;
  \item This pattern suggests that the highest transcendental function is a
    sequence of $0$ and $-1$.
  \end{itemize}
The results obtained in this way reproduce the previous ones, in particular (remember that the loop order is equivalent to $\kappa-1$)
\begin{align}
\begin{cases}
\frac{4^{ \kappa} \kappa !}{15 \cdot 32}  i^{\kappa} \frac{1}{x^3(1-x)^3}p_{\kappa}\left(\sfrac{-1}{x}\right) & \text{for } \kappa\text{ even} \\
\frac{4^{ \kappa} \kappa !}{15 \cdot 32}  i^{\kappa +1} \frac{1}{x^3(1-x)^3}p_{\kappa}\left(\sfrac{1}{1+x}\right) & \text{for } \kappa\text{ odd} 
\end{cases}
\end{align}
We can play a similar game for the next transcendental piece, in this
  case the iterative structure is a bit more involved:
  \begin{itemize}
    \item both $\mathcal{K}^{\pm}(v,x)$ functions will contribute at
      each loop order;
    \item contrary to what happens before, now we have two contributions,
      one coming from the highest transcendental polynomial and one
      coming from the next to the highest  at the previous loop;
      \item for the highest transcendental polynomial at even loop we
        have to obtain the general patter for the integration of
        \begin{equation}
         \frac{ H_{0,\ldots}(v-1)}{v^i} \qquad  H_{0,\ldots}(v-1)  v^i \,,
       \end{equation}
       for both $\mathcal{K}^{\pm}(v,x)$ functions in the associated integration regions $\{0,1+x\}$ and $\{1+x,1\}$.
    \item for odd loops we have to do the same but for a different set
      of functions and pole structure, namely:
      \begin{equation}
         \frac{ H_{-1,\ldots}(v-1)}{(v-1)^i} \qquad  H_{-1,\ldots}(v-1)  v^i \,.
       \end{equation}
    \item The next to the highest transcendental piece at the previous order will contribute only when
      integrated against a simple pole in $\{v,(v-1)\}$. The pattern
      is similar to the one obtained for the highest transcendental
      polynomials just in this case the poles contributions are
      inverted\footnote{This happens because contrary to the highest
        transcendental part where only $x$ and $(1+x)$ can appear in the
        denominator at each loop for the next to the highest one both
        contributions appear.}.
    \end{itemize}
    As before we can express the results in terms of the $q_{\kappa}$'s:
\begin{align}
\begin{cases}
 \frac{-4^{ \kappa} \kappa !}{15 \cdot 32}  i^{\kappa} \frac{1}{x^3(1-x)^3}q_{\kappa}\left(\sfrac{1}{1+x}\right) & \text{for } \kappa\text{ even} \\
\frac{4^{ \kappa} \kappa !}{15 \cdot 32}  i^{\kappa +1} \frac{1}{x^3(1-x)^3}q_{\kappa}\left(\sfrac{-1}{x}\right)&  \text{for } \kappa\text{ odd} 
\end{cases}
\end{align}
We notice  that, as already in Eq.~\eqref{pandqPol}, the arguments of the $p$'s and $q$'s are alternating. In addition, if we translate these results in terms of $z=\frac{1}{1+x}$ we notice that the polynomials that we find from the amplitude correspond to the one multiplying $I_{\kappa}$ and $I_{\kappa-1}$. Interestingly these integrals are the ones producing $\log^2 V$, confirming the connection between CFT logarithmic singularities and amplitudes, as we have  explored in Mellin space in Sec.~\ref{sec:Mell2l}. 
\bibliographystyle{JHEP}
\bibliography{biblo}

\end{document}